\documentclass[11pt]{article}
\usepackage{pst-plot,pst-node}
\usepackage{latexsym}
\usepackage{amssymb}
\usepackage{epsfig}
\usepackage{amsfonts}
\usepackage{color}
\usepackage{amsmath}
\begin{document}

\newcommand{\bull}{\rule{.85ex}{1ex} \par \bigskip}
\newenvironment{sketch}{\noindent {\bf Proof (sketch):\ }}{\hfill \bull}
\newenvironment{example}{\begin{exmp} \rm }{\hfill $\Box$ \end{exmp}}

\newtheorem{theorem}{Theorem}[section]
\newtheorem{definition}[theorem]{Definition}
\newtheorem{proposition}[theorem]{Proposition}
\newtheorem{lemma}[theorem]{Lemma}
\newtheorem{corollary}[theorem]{Corollary}
\newtheorem{conjecture}[theorem]{Conjecture}
\newtheorem{exmp}[theorem]{Example}
\newtheorem{notation}[theorem]{Notation}
\newtheorem{problem}{Problem}
\newtheorem{remark}[theorem]{Remark}
\newtheorem{observation}[theorem]{Observation}

\newcommand{\QCSP}[1]{\mbox{\rm QCSP$(#1)$}}
\newcommand{\CSP}[1]{\mbox{\rm CSP$(#1)$}}
\newcommand{\MCSP}[1]{\mbox{{\sc Max CSP}$(#1)$}}
\newcommand{\wMCSP}[1]{\mbox{\rm weighted Max CSP$(#1)$}}
\newcommand{\cMCSP}[1]{\mbox{\rm cw-Max CSP$(#1)$}}
\newcommand{\tMCSP}[1]{\mbox{\rm tw-Max CSP$(#1)$}}
\renewcommand{\P}{\mbox{\bf P}}
\newcommand{\G}[1]{\mbox{\rm I$(#1)$}}
\newcommand{\NE}[1]{\mbox{$\neq_{#1}$}}

\newcommand{\NP}{\mbox{\bf NP}}
\newcommand{\NL}{\mbox{\bf NL}}
\newcommand{\PO}{\mbox{\bf PO}}
\newcommand{\NPO}{\mbox{\bf NPO}}
\newcommand{\APX}{\mbox{\bf APX}}

\newcommand{\GIF}[3]{\ensuremath{h_\{{#2},{#3}\}^{#1}}}

\newcommand{\Spmod}{\mbox{\rm Spmod}}
\newcommand{\Sbmod}{\mbox{\rm Sbmod}}

\newcommand{\Inv}[1]{\mbox{\rm Inv($#1$)}}
\newcommand{\Pol}[1]{\mbox{\rm Pol($#1$)}}
\newcommand{\sPol}[1]{\mbox{\rm s-Pol($#1$)}}

\newcommand{\un}{\underline}
\newcommand{\ov}{\overline}
\def\ar{\hbox{ar}}
\def\vect#1#2{#1 _1\zdots #1 _{#2}}
\def\zd{,\ldots,}
\let\sse=\subseteq
\let\la=\langle
\def\lla{\langle\langle}
\let\ra=\rangle
\def\rra{\rangle\rangle}
\let\vr=\varrho
\def\vct#1#2{#1 _1\zd #1 _{#2}}
\newcommand{\va}{{\bf a}}
\newcommand{\vb}{{\bf b}}
\newcommand{\vc}{{\bf c}}
\newcommand{\bx}{{\bf x}}
\newcommand{\by}{{\bf y}}
\def\Z{{\bur Z^+}}
\def\R{{\bur R}}
\def\D{{\cal D}}
\def\F{{\cal F}}
\def\I{{\cal I}}
\def\C{{\cal C}}
\def\U{{\cal U}}
\def\K{{\cal K}}
\def\Lat{{\cal L}}

\def\2mat#1#2#3#4#5#6#7#8{
\begin{array}{c|cc}
$~$ & #3 & #4\\
\hline
#1 & #5& #6\\
#2 & #7 & #8 \end{array}}

\font\tenbur=msbm10
\font\eightbur=msbm8
\def\bur{\fam11}
\textfont11=\tenbur \scriptfont11=\eightbur \scriptscriptfont11=\eightbur
\font\twelvebur=msbm10 scaled 1200
\textfont13=\twelvebur \scriptfont13=\tenbur \scriptscriptfont13=\eightbur




\newcounter{A}
\newcounter{prgline} 


\newcommand{\citelist}[1]{\raisebox{.2ex}{[}#1\raisebox{.2ex}{]}}
\newcommand{\scite}[1]{\citeauthor{#1}, \citeyear{#1}}
\newcommand{\shortcite}[1]{\cite{#1}}
\newcommand{\mciteii}[2]{\citeauthor{#1}, \citeyear{#1}, %
\citeyear{#2}}
\newcommand{\mciteiii}[3]{\citeauthor{#1}, \citeyear{#1}, %
\citeyear{#2}, \citeyear{#3}}

\newcommand{\multiciteii}[2]{\citelist{\scite{#1}, \citeyear{#2}}}
\newcommand{\multiciteiii}[3]%
  {\citelist{\scite{#1}, \citeyear{#2}, \citeyear{#3}}}


\renewcommand{\phi}{\varphi}
\renewcommand{\epsilon}{\varepsilon}

\newcommand{\draft}{\begin{center}\huge Draft!!! \end{center}}
\newcommand{\void}{\makebox[0mm]{}}     


\renewcommand{\text}[1]{\mbox{\rm \,#1\,}}        

\newcommand{\tand}{\text{\ and\ }}
\newcommand{\tor}{\text{\ or\ }}
\newcommand{\tif}{\text{\ if\ }}
\newcommand{\tiff}{\text{\ iff\ }}
\newcommand{\tfor}{\text{\ for\ }}
\newcommand{\tforall}{\text{\ for all\ }}
\newcommand{\totherwise}{\text{\ otherwise}}


\newcommand{\fnl}{\void \\}             

\newcommand{\pushlist}[1]{\setcounter{#1}{\value{enumi}} \end{enumerate}}
\newcommand{\poplist}[1]{\begin{enumerate} \setcounter{enumi}{\value{#1}}}


\renewcommand{\emptyset}{\varnothing}  
\newcommand{\union}{\cup}               
\newcommand{\intersect}{\cap}           
\newcommand{\setdiff}{-}                
\newcommand{\compl}[1]{\overline{#1}}   
\newcommand{\card}[1]{{|#1|}}           
\newcommand{\set}[1]{\{{#1}\}} 
\newcommand{\st}{\ |\ }                 
\newcommand{\suchthat}{\st}             
\newcommand{\cprod}{\times}             
\newcommand{\powerset}[1]{{\bf 2}^{#1}} 

\newcommand{\tuple}[1]{\langle{#1}\rangle}  
\newcommand{\seq}[1]{\langle #1 \rangle}
\newcommand{\emptyseq}{\seq{}}
\newcommand{\floor}[1]{\left\lfloor{#1}\right\rfloor}
\newcommand{\ceiling}[1]{\left\lceil{#1}\right\rceil}

\newcommand{\map}{\rightarrow}
\newcommand{\fncomp}{\!\circ\!}         

\newcommand{\transclos}[1]{#1^+}
\newcommand{\reduction}[1]{#1^-}        

\newtheorem{defnx}{Definition}
\newtheorem{axiomx}[defnx]{Axiom}
\newtheorem{theoremx}[defnx]{Theorem}
\newtheorem{propositionx}[defnx]{Proposition}
\newtheorem{lemmax}[defnx]{Lemma}
\newtheorem{corollx}[defnx]{Corollary}
\newtheorem{algx}[defnx]{Algorithm}
\newtheorem{exx}[defnx]{Example}
\newtheorem{factx}[defnx]{Fact}

\newcommand{\QED}{\nopagebreak[4]{\makebox[1mm]{}\hfill$\Box$}}
\newenvironment{proof}{\noindent {\bf \noindent Proof: }}{\QED \\}
\newenvironment{proofsk}{\noindent {\bf \noindent Proof sketch: }}{\QED
\vspace{-\baselineskip}}


\newlength{\prgindent}
\newenvironment{program}{
  \begin{list}%
   {\arabic{prgline}}%
   {
   \usecounter{prgline}
   \setlength{\prgindent}{0em}
   \setlength{\parsep}{0em}
   \setlength{\itemsep}{0em}
   \setlength{\labelwidth}{1em}
   \setlength{\labelsep}{1em}
   \setlength{\leftmargin}{\labelwidth}
   \addtolength{\leftmargin}{\labelsep}
   \setlength{\topsep}{0em}
   \setlength{\parskip}{0em} } }%
 {\end{list}}

\newif\ifprgendtext
\prgendtexttrue

\newenvironment{prgblock}{\addtolength{\prgindent}{\labelsep}}%
{\addtolength{\prgindent}{-\labelsep}}
\newcommand{\prgbeginblock}{\addtolength{\prgindent}{\labelsep}}
\newcommand{\prgendblock}{\addtolength{\prgindent}{-\labelsep}}
\newcommand{\prgcndendblock}[1]{\addtolength{\prgindent}{-\labelsep}
 \ifprgendtext \prglin\prgres{#1}\fi}

\newcommand{\prglin}{\rm \item\hspace{\prgindent}}
\newcommand{\prgcontlin}{\\  \hspace{\prgindent}}

\newcommand{\prgres}[1]{{\bf #1}}
\newcommand{\prgassn}{\leftarrow}
\newcommand{\prgname}[1]{{\it #1}}

\newcommand{\prgbegin}{\prglin\prgres{begin}\prgbeginblock}
\newcommand{\prgend}{\prgendblock\prglin\prgres{end\ }}
\newcommand{\prgnoend}{\prgendblock}

\newcommand{\prgif}{\prglin\prgres{if\ }}
\newcommand{\prgthen}{\prgres{\ then\ }\prgbeginblock}
\newcommand{\prgelse}{\prgendblock\prglin\prgres{else}\prgbeginblock}
\newcommand{\prgelsif}{\prgendblock\prglin\prgres{elsif\ }}
\newcommand{\prgendif}{\prgcndendblock{end if}}

\newcommand{\prgwhile}{\prglin\prgres{while\ }}
\newcommand{\prgfor}{\prglin\prgres{for\ }}
\newcommand{\prgdo}{\prgres{\ do}\prgbeginblock}
\newcommand{\prgrepeat}{\prgres{\ repeat}\prgbeginblock}
\newcommand{\prgloop}{\prglin\prgres{loop}\prgbeginblock}
\newcommand{\prgendloop}{\prgendblock\prglin\prgres{end loop}}
\newcommand{\prgendwhile}{\prgcndendblock{end while}}
\newcommand{\prgendfor}{\prgcndendblock{end for}}
\newcommand{\prguntil}{\prgendblock\prglin\prgres{until\ }}

\newcommand{\prgcomment}{\prglin\prgres{comment\ }\it }
\newcommand{\prgprocedure}{\prglin\prgres{procedure\ }}
\newcommand{\prgnil}{\prgres{\ nil}}
\newcommand{\prgtrue}{\prgres{\ true}}
\newcommand{\prgfalse}{\prgres{\ false}}
\newcommand{\prgnot}{\prgres{\ not\ }}
\newcommand{\prgand}{\prgres{\ and\ }}
\newcommand{\prgor}{\prgres{\ or \ }}
\newcommand{\prgfail}{\prgres{fail}}
\newcommand{\prgreturn}{\prgres{return\ }}
\newcommand{\prgaccept}{\prgres{accept}}
\newcommand{\prgreject}{\prgres{reject}}

\prgendtextfalse

\newcommand{\note}[1]{{\tt #1}}


\newcommand{\ie}{{\em ie.}}                
\newcommand{\eg}{{\em eg.}}
\newcommand{\paper}{paper}                

\newcommand{\emdef}{\em}                   
\newcommand{\rinterpretation}{${\Bbb R}$-interpretation}
\newcommand{\rmodel}{${\Bbb R}$-model}
\newcommand{\transp}{^{\rm T}}

\newcommand{\unprint}[1]{}
\newcommand{\blankline}{$\:$}

\newcommand{\Solv}{{\it TSolve}}
\newcommand{\Neg}{{\it Neg}}
\newcommand{\logname}{XX}

\newcommand{\props}{{\it props}}
\newcommand{\rels}{{\it rels}}
\newcommand{\deduce}{\vdash_p}

\newcommand{\pform}{{\rm Pr}}
\newcommand{\axform}{{\rm AX}}
\newcommand{\axset}{{\bf AX}}
\newcommand{\resdeduce}{\vdash_{\rm R}}
\newcommand{\resaxdeduce}{\vdash_{\rm R,A}}

\newcommand{\cmis}{{\em \#mis}}
\newcommand{\combine}{{\em comb}}

\newcommand{\xcsp}{{\sc X-Csp}}
\newcommand{\csp}{{\sc Csp}}

\author{
Vladimir Deineko\\
Warwick Business School\\
University of Warwick, UK\\
$\texttt{Vladimir.Deineko@wbs.ac.uk}$\\
\and
Peter Jonsson\\
Dep't of Computer and Information Science\\
University of Link\"oping, Sweden\\
$\texttt{peter.jonsson@ida.liu.se}$\\
\and
Mikael Klasson\\
Dep't of Computer and Information Science\\
University of Link\"oping, Sweden\\
$\texttt{mikael.klasson@ida.liu.se}$\\
\and
Andrei Krokhin\\
Department of Computer Science\\
University of Durham, UK\\
$\texttt{andrei.krokhin@durham.ac.uk}$\\
}
\title{The approximability of {\sc Max CSP} with fixed-value constraints}

\date{}
\maketitle
\bibliographystyle{plain}


\begin{abstract}
In the maximum constraint satisfaction problem ({\sc Max CSP}),
one is given a finite collection of (possibly weighted)
constraints on overlapping sets of variables, and the goal is to
assign values from a given finite domain to the variables so as to
maximize the number (or the total weight, for the weighted case)
of satisfied constraints. This problem is \NP-hard in general,
and, therefore, it is natural to study how restricting the allowed
types of constraints affects the approximability of the problem.
In this paper, we show that  any {\sc Max CSP} problem with a
finite set of allowed constraint types, which includes all
fixed-value constraints (i.e., constraints of the form $x=a$), is
either solvable exactly in polynomial time or else is
$\APX$-complete, even if the number of occurrences of variables in
instances is bounded. Moreover, we present a simple description of
all polynomial-time solvable cases of our problem. This
description relies on the well-known algebraic combinatorial
property of supermodularity.
\end{abstract}

\medskip


\bigskip

\section{Introduction and Related Work}

\subsection{Background}

Many combinatorial optimization problems are $\NP$-hard, and the
use of approximation algorithms is one of the most prolific
techniques to deal with $\NP$-hardness. However, hard optimization
problems exhibit different behaviour with respect to
approximability, and complexity theory for approximation is now a
well-developed area~\cite{Ausiello99:complexity}.

Constraint satisfaction problems (CSPs) have always played a
central role in this direction of research, since the CSP
framework contains many natural computational problems, for
example, from propositional logic and graph theory (see,
e.g.,~\cite{Creignouetal:siam01,Hell04:book}).
In a CSP, informally speaking, one is given a finite collection of
constraints on overlapping sets of variables, and the goal is to
decide whether there is an assignment of values from a given
domain to the variables satisfying all constraints (decision
problem) or to find an assignment satisfying maximum number of
constraints (optimization problem). These are the main versions of
the CSP, and there are many other versions obtained from them by
modifying the objective (see,
e.g.,~\cite{Creignouetal:siam01,Krokhin03:functions,Krokhin04:survey}).
In this paper, we will focus on the optimization problems, which
are known as {\em maximum constraint satisfaction} problems, {\sc
Max CSP} for short. The most well-known examples of such problems
are {\sc Max $k$-Sat} and {\sc Max Cut}. Let us now formally
define {\sc Max CSP}.

Let $D$ denote a {\em finite} set with $|D|>1$. Let $R^{(m)}_D$
denote the set of all $m$-ary predicates over $D$, that is,
functions from $D^m$ to $\{0,1\}$, and let
$R_D=\bigcup_{m=1}^{\infty} R^{(m)}_D$. Also, let $\Z$ denote the
set of all non-negative integers.

\begin{definition}[constraint]
A {\em constraint} over a set of variables
$V=\{x_1,x_2,\ldots,x_n\}$ is an expression of the form $f({\bf
x})$ where

\begin{itemize}
\item $f \in R^{(m)}_D$ is called the {\em constraint predicate};
and

\item ${\bf x} = (x_{i_1},\ldots,x_{i_m})$ is called the {\em
constraint scope}.
\end{itemize}

The constraint $f({\bf x})$ is said to be {\em satisfied} on a
tuple ${\bf a}=(a_{i_1},\ldots,a_{i_m}) \in D^m$ if $f({\bf
a})=1$.
\end{definition}

Note that throughout the paper the values 0 and 1 taken by any
predicate will be considered as integers, not as Boolean values,
and addition will always denote the addition of integers.

\begin{definition}[{\sc Max CSP}]
For a finite $\F\sse R_D$, an instance of $\MCSP\F$ is a pair
$(V,C)$ where
\begin{itemize}
\item $V=\{x_1,\ldots,x_n\}$ is a set of variables taking their
values from the set $D$;

\item $C$ is a collection of constraints $f_1({\bf
x}_1),\ldots,f_q({\bf x}_q)$ over $V$, where $f_i \in \F$ for all
$1 \leq i \leq q$.
\end{itemize}
The goal is to find an assignment $\phi:V\rightarrow D$ that
maximizes the number of satisfied constraints, that is, to
maximize the function $f:D^n \rightarrow \Z$, defined by
$f(x_1,\ldots,x_n)=\sum_{i=1}^q f_i({\bf x}_i)$. If the
constraints have (positive integral) weights $\varrho_i$, $1\le
i\le q$, then the goal is to maximize the total weight of
satisfied constraints, that is, to maximize the function $f:D^n
\rightarrow \Z$, defined by $f(x_1,\ldots,x_n)=\sum_{i=1}^q
\varrho_i\cdot f_i({\bf x}_i)$.
\end{definition}

Complexity classifications for various versions of constraint
satisfaction problems have attracted much attention in the recent
years (see surveys~\cite{Krokhin03:functions,Krokhin04:survey})
because, as the authors of~\cite{Creignouetal:siam01} nicely put
it, these classifications ``present as reasonably accurate bird's
eye view of computational complexity and the equivalence classes
it has created''. Classifications with respect to a set of allowed
constraint types (such as $\F$ in $\MCSP\F$ above) have been of
particular interest, e.g.,\cite{Boerner03:quantified,Bulatov03:conservative,%
Bulatov03:counting,Bulatov??:classifying,%
Creignouetal:siam01,Feder98:monotone,Hell03:algorithmic}.

{\em Boolean} constraint satisfaction problems (that is, when
$D=\{0,1\}$) are by far better studied~\cite{Creignouetal:siam01}
than the non-Boolean version. The main reason is, in our opinion,
that Boolean constraints can be conveniently described by
propositional formulas which provide a flexible and easily
manageable tool, and which have been extensively used in
complexity theory from its very birth. Moreover, Boolean CSPs
suffice to represent a number of well-known problems and to obtain
results clarifying the structure of complexity for large classes
of interesting problems~\cite{Creignouetal:siam01}. In particular,
Boolean CSPs were used to provide evidence for one of the most
interesting phenomena in complexity theory, namely that
interesting problems belong to a small number of complexity
classes~\cite{Creignouetal:siam01}, which cannot be taken for
granted due to Ladner's theorem. After the pioneering work of
Schaefer~\cite{Schaefer78:complexity} presenting a tractable
versus $\NP$-complete dichotomy for Boolean decision CSPs, many
classification results have been obtained (see,
e.g.,~\cite{Creignouetal:siam01}), most of which are dichotomies.
In particular, a dichotomy in complexity and approximability for
Boolean {\sc Max CSP} has been obtained by
Creignou~\cite{Creignou95:maximum}, and it was slightly refined
in~\cite{Khanna01:approximability} (see
also~\cite{Creignouetal:siam01}). The complexity of Boolean {\sc
Max CSP} with arbitrary (i.e., not necessarily positive) weights
was classified in~\cite{Jonsson00:boolean}.

Many papers on various versions of Boolean CSPs mention studying
non-Boolean CSPs as a possible direction of future research, and
additional motivation for it, with an extensive discussion, was
given by Feder and Vardi~\cite{Feder98:monotone}. Dichotomy
results on non-Boolean CSPs give a better understanding of what
makes a computational problem tractable or hard, and they give a
more clear picture of the structure of complexity of problems,
since many facts observed in Boolean CSPs appear to be special
cases of more general phenomena. Notably, many appropriate tools
for studying non-Boolean CSPs have not been discovered until
recently. For example, universal algebra tools have proved to be
very fruitful when working with decision, counting, and quantified
CSPs~\cite{Boerner03:quantified,Bulatov02:dichotomy,Bulatov03:conservative,Bulatov03:counting,Bulatov??:classifying}
while ideas from lattice theory, combinatorial optimization and
operations research have been recently suggested for optimization
problems~\cite{Cohen05:supermodular,Krokhin05:diamonds}.

The problem {\sc Max CSP} is $\NP$-hard in general (i.e., without
restrictions on the type of allowed constraints), and there is a
significant body of results on algorithmic and
complexity-theoretical aspects of this problem, including results
on superpolynomial general algorithms
(e.g.,\cite{Datar03:combinatorial,Williams04:optimal}), polynomial
algorithms for special
cases~\cite{Cohen05:supermodular,Krokhin05:diamonds}, explicit
approximability bounds
(e.g.,~\cite{Engebretsen04:non-approx,Hast05:thesis,Hastad01:optimal,Hastad05:2csp,Khot04:optimal}),
and complexity of approximation (e.g., ~\cite{Bazgan05:global,%
Creignouetal:siam01,Jonsson04:maxcsp3}).

The main research problem that we will look at in this paper is
the following.

\begin{problem}\label{problem}
Classify the problems $\MCSP\F$ with respect to approximability.
\end{problem}

We say that a predicate is {\em non-trivial} if it is not
identically 0. We will always assume that $\F$ is finite and
contains only non-trivial predicates. Whenever we do not specify
which version (weighted or unweighted) we consider, we mean {\em
unweighted} {\sc Max CSP}. Note that the definition allows one to
repeat constraints in instances (we
follow~\cite{Creignouetal:siam01} in this), so our unweighted
problem actually allows polynomially bounded weights. However, our
tractability results will hold for the weighted version, while in
our hardness results, for every $\F$, we will use only instances
where every constraint occurs at most $k_\F$ times (where $k_\F$
is a constant depending on $\F$).

For the Boolean case, Problem~\ref{problem} was solved
in~\cite{Creignou95:maximum,Creignouetal:siam01,Khanna01:approximability}.
It appears that Boolean $\MCSP\F$ problems exhibit a dichotomy in
that such a problem is either solvable exactly in polynomial time
or else $\APX$-complete, i.e., does not admit a PTAS
(polynomial-time approximation scheme) unless \P=\NP. These papers
also describe the boundary between the two cases. This dichotomy
result was extended to the case $|D|=3$
in~\cite{Jonsson04:maxcsp3}, which is to the best of our knowledge
the only paper tackling Problem~\ref{problem} in the non-Boolean
case.

\subsection{Results}

For a subset $D'\sse D$, let $u_{D'}$ denote the predicate such
that $u_{D'}(x)=1$ if and only if $x \in D'$. Let ${\cal
U}_D=\{u_{D'} \; | \; \emptyset\ne D' \subseteq D\}$, that is,
$\U_D$ is the set of all non-trivial unary predicates on $D$.
Furthermore, let $\C_D=\{u_{\{d\}}\mid d\in D\}$. Note that
predicates from ${\cal C}_D$ give rise to constraints of the form
$x=d$, i.e., fixed-value constraints.

The decision problems $\CSP\F$ are similar to $\MCSP\F$, but the
the task is to decide whether all constraints in a given instance
can be simultaneously satisfied. Problems of the form $\CSP{\F\cup
\U_D}$ are known as conservative (or list) CSPs, and their
complexity has been completely classified by Bulatov
in~\cite{Bulatov03:conservative}, while a complexity
classification for the problems of the form $\CSP{\F\cup \C_D}$
would imply a classification for all problems
$\CSP\F$~\cite{Bulatov??:classifying}.

In this paper we solve the above Problem~\ref{problem} for all
sets of the form $\F\cup\C_D$ where $D$ is any finite set. (Note
that this does not necessarily imply a full solution to
Problem~\ref{problem}, as it would for decision problems.) Our
result is parallel to Bulatov's classification of conservative
CSPs~\cite{Bulatov03:conservative}, but our techniques are quite
different from the universal-algebraic techniques used
in~\cite{Bulatov03:conservative}. The universal-algebraic
techniques
from~\cite{Bulatov03:conservative,Bulatov??:classifying} cannot be
applied in the optimization setting because the basic properties
of decision CSPs that make these techniques useful are not
satisfied by {\sc Max CSP}.

It was suggested in Section 6 of~\cite{Cohen05:supermodular} that
$\MCSP{\F\cup\C_D}$ is solvable exactly in polynomial time if and
only if all predicates in $\F$ are supermodular with respect to
some linear ordering on $D$ (see definitions in
Section~\ref{supmoddefsection}). We prove that this is indeed the
case, and that in all other cases the problem $\MCSP{\F\cup\C_D}$
is $\APX$-complete. Moreover, we show that every $\APX$-complete
problem of the above form is $\APX$-complete even when we further
restrict it to instances where the number of occurrences of
variables is bounded by some (possibly large) constant. Note that
approximability properties for constraint problems with the
bounded occurrence property (as well as for related problems on
graphs with bounded degree) have been intensively studied in the
literature (see,
e.g.,~\cite{Alimonti00:cubic,Berman03:small,Hastad00:bounded,Karpinski01:bounded}).

Our classification result uses the combinatorial property of
supermodularity which is a well-known source of tractable
optimization
problems~\cite{Burkard96:Monge,Fujishige05:submodular,Topkis98:book},
and the technique of strict
implementations~\cite{Creignouetal:siam01,Khanna01:approximability}
which allows one to show that an infinite family of problems can
express, in a regular way, one of a few basic hard problems. We
remark that the idea to use supermodularity in the analysis of the
complexity of $\MCSP\F$ is very new, and has not been even
suggested in the literature prior to~\cite{Cohen05:supermodular}.
It was shown in~\cite{Cohen05:supermodular,Jonsson04:maxcsp3} that
supermodularity is the {\em only} source of tractability for
problems of the form $\MCSP{\F}$ when $D$ is small (i.e., $|D|\le
3$). This, together with the results obtained in the present
paper, suggests that supermodularity is indeed the appropriate
tool for tackling Problem~\ref{problem}.

Some of our technical results (those in Appendix A) are of
independent interest in combinatorics.
In~\cite{Klinz95:permuting}, Klinz {\em et al.} study how one can
permute rows and columns of a 0-1 matrix so as to avoid a
collection of given forbidden submatrices; some results of this
nature have later been used in constructing phylogenetic
trees~\cite{Pe'er04:incomplete}. Klinz {\em et al.} obtain many
results in this direction, but they leave open the case when
matrices are square and rows and columns must be permuted by the
{\em same} permutation (see Section 6
of~\cite{Klinz95:permuting}). Our results clarify the situation in
this special case for one type of forbidden matrices considered in
Theorem 4.5 of~\cite{Klinz95:permuting}.



The structure of the paper is as follows: Section~\ref{basics}
contains definitions of approximation complexity classes and
reductions. In Section~\ref{techniques}, we describe our reduction
techniques, and in Section~\ref{supmoddefsection} we give the
basics of supermodularity and  discuss the relevance of
supermodularity in the study of {\sc Max CSP}.
Section~\ref{mainsec} contains the proof of the main theorem of
the paper. Finally, In Section~\ref{hcolsec}, we discuss an
application of our results to the optimization version of the {\sc
List $H$-colouring} problem for digraphs. 
Some of the technical proofs omitted from the main body of the
paper can be found in Appendices.

\section{Basics of approximability}\label{basics}

A {\em combinatorial optimization problem} is defined over a set
of {\em instances} (admissible input data); each instance $\I$ has
a finite set ${\sf sol}(\I)$ of {\em feasible solutions}
associated with it. The {\em objective function} attributes a
positive integer cost to every solution in ${\sf sol}(\I)$. The
goal in an optimization problem is, given an instance $\I$, to
find a feasible solution of {\em optimum} cost. The optimal cost
is the largest one for {\em maximization} problems and the
smallest one for {\em minimization} problems. A combinatorial
optimization problem is said to be an $\NP$ optimization ($\NPO$)
problem if its instances and solutions can be recognized in
polynomial time, the solutions are polynomial-bounded in the input
size, and the objective function can be computed in polynomial
time (see, e.g.,~\cite{Ausiello99:complexity}).

\begin{definition}[performance ratio]
A solution $s\in {\sf sol}(\I)$ to an instance $\mathcal{I}$ of an
{\bf NPO} problem $\Pi$ is $r$-approximate if
$$
\max{\{ \frac{cost(s)}{Opt(\I)},\frac{Opt(\I)}{cost(s)} \} }\le r,
$$
where $Opt(\I)$ is the optimal cost for a solution to $\I$. An
approximation algorithm for an $\NPO$ problem $\Pi$ has {\em
performance ratio} $\mathcal{R}(n)$ if, given any instance $\I$ of
$\Pi$ with $|\I|=n$, it outputs an $\mathcal{R}(n)$-approximate
solution.
\end{definition}

\begin{definition}[complexity classes]
{\bf PO} is the class of {\bf NPO} problems that can be solved (to
optimality) in polynomial time. An {\bf NPO} problem $\Pi$ is in
the class $\APX$ if there is a polynomial time approximation
algorithm for $\Pi$ whose performance ratio is bounded by a
constant.
\end{definition}

The following result is contained in
Proposition~2.3~\cite{Cohen05:supermodular} and its proof.

\begin{lemma}\label{genapprox}
Every (weighted or not) problem $\MCSP\F$ belongs to $\APX$.
Moreover, if $a$ is the maximum arity of any predicate in $\F$
then there is a polynomial time algorithm which, for every
instance $\I$ of $\MCSP\F$, produces a solution satisfying at
least $\frac{q}{|D|^a}$ constraints, where $q$ is the number of
constraints in $\I$.
\end{lemma}

Completeness in $\APX$ is defined using an appropriate reduction,
called $AP$-reduction. Our definition of this reduction
follows~\cite{Creignouetal:siam01,Khanna01:approximability}.

\begin{definition}[$AP$-reduction, $\APX$-completeness]
An $\NPO$ problem $\Pi_1$ is said to be {\em $AP$-reducible} to an
$\NPO$ problem $\Pi_2$ if two polynomial-time computable functions
$F$ and $G$ and a constant $\alpha$ exist such that
\begin{itemize}
\item[(a)] for any instance $\I$ of $\Pi_1$, $F(\I)$ is an
instance of $\Pi_2$;

\item[(b)] for any instance $\I$ of $\Pi_1$, and any feasible
solution $s'$ of $F(\I)$, $G(\I,s')$ is a feasible solution of
$\I$;

\item[(c)] for any instance $\I$ of $\Pi_1$, and any $r\ge 1$, if
$s'$ is an $r$-approximate solution of $F(\I)$ then $G(\I,s')$ is
an $(1+(r-1)\alpha+o(1))$-approximate solution of $\I$ where the
$o$-notation is with respect to $|\I|$.
\end{itemize}

An $\NPO$ problem $\Pi$ is {\em $\APX$-hard} if every problem in
$\APX$ is $AP$-reducible to it. If, in addition, $\Pi$ is in
$\APX$ then $\Pi$ is called {\em $\APX$-complete}.
\end{definition}

It is a well-known fact (see, e.g., Section 8.2.1
in~\cite{Ausiello99:complexity}) that $AP$-reductions compose.  We
shall now give an example of an $\APX$-complete problem which will
be used extensively in this paper.

\begin{exmp}\label{maxkcol}
Given a graph $G=(V,E)$, the {\sc Maximum $k$-colourable Subgraph}
problem, $k\ge 2$, is the problem of maximizing $|E'|$, $E'\sse
E$, such that the graph $G'=(V,E')$ is $k$-colourable. This
problem is known to be \APX-complete (it is Problem GT33
in~\cite{Ausiello99:complexity}). Let $neq_k$ denote the binary
disequality predicate on $D=\{0,1\zd k-1\}$, $k\ge 2$, that is,
$neq_k(x,y)=1 \Leftrightarrow x\ne y$. Consider the problem
$\MCSP{\{neq_k\}}$ restricted to instances where every pair of
variables appears in the scope of at most one constraint. This
problem is exactly the {\sc Maximum $k$-colourable Subgraph}
problem. To see this, think of vertices of a given graph as of
variables that take values from $D$, and introduce the constraint
$neq_k(x,y)$ for every pair of variables $x,y$ such that $(x,y)$
is an edge in the graph. It follows that the problem
$\MCSP{\{neq_k\}}$ is \APX-complete.

Note that the weighted $\MCSP{\{neq_k\}}$ problem coincides with
the well-known problem {\sc Max $k$-Cut} (it is Problem ND17
in~\cite{Ausiello99:complexity}). The {\sc Max 2-Cut} problem is
usually referred to as simply {\sc Max Cut}.
\end{exmp}


In some of our hardness proofs, it will be convenient for us to
use another type of approximation-preserving reduction, called an
$L$-reduction~\cite{Ausiello99:complexity}.

\begin{definition}[$L$-reduction]
An $\NPO$ problem $\Pi_1$ is said to be {\em $L$-reducible} to an
$\NPO$ problem $\Pi_2$ if two polynomial-time computable functions
$F$ and $G$ and positive constants $\alpha,\beta$ exist such that
\begin{itemize}
\item[(a)] given any instance $\I$ of $\Pi_1$, algorithm $F$
produces an instance $\I'=F(I)$ of $\Pi_2$, such that the cost of
an optimal solution for $\I'$, $Opt(\I')$, is at most $\alpha\cdot
Opt(\I)$;

\item[(b)] given $\I, \I'=F(\I)$, and any solution $s'$ to $\I'$,
algorithm $G$ produces a solution $s$ to $\I$ such that
$|cost(s)-OPT(\I)|\le \beta\cdot |cost(s')-OPT(\I')|$.
\end{itemize}
\end{definition}
It is well known (see, e.g., Lemma 8.2
in~\cite{Ausiello99:complexity}) that, within $\APX$, the
existence of an $L$-reduction from $\Pi_1$ to $\Pi_2$ implies the
existence of an $AP$-reduction from $\Pi_1$ to $\Pi_2$.

\section{Reduction techniques}\label{techniques}

The basic reduction technique in our $\APX$-completeness proofs is
based on {\em strict implementations},
see~\cite{Creignouetal:siam01,Khanna01:approximability} where this
notion was introduced for the Boolean case. We will give this
definition in a slightly different form from that
of~\cite{Creignouetal:siam01,Khanna01:approximability}, but it can
easily be checked to be equivalent to the original one (in the
case $|D|=2$).

\begin{definition}[strict implementation]\label{strictdef}
Let $Y=\{y_1,\ldots,y_m\}$ and $Z=\{z_1,\ldots,z_n\}$ be two
disjoint sets of variables. The variables in $Y$ are called
primary and the variables in $Z$ auxiliary. The set $Z$ may be
empty. Let $g_1({\bf y}_1),\ldots,g_s({\bf y}_s)$, $s > 0$, be
constraints over $Y \cup Z$. If $g(y_1,\ldots,y_m)$ is a predicate
such that the equality
$$
g(y_1,\ldots,y_m) + (\alpha -1)= \max_{Z}{\sum_{i=1}^{s}{g_i({\bf
y}_i)}}
$$
is satisfied for all $y_1,\ldots,y_m$, and some fixed $\alpha \in
\Z$, then this equality is said to be a {\em strict
\mbox{$\alpha$-implementation}} of $g$ from $g_1\zd g_s$.

\end{definition}
We use $\alpha-1$ rather than $\alpha$ in the above equality to
ensure that this notion coincides with the original notion of a
strict $\alpha$-implementation for Boolean
constraints~\cite{Creignouetal:siam01,Khanna01:approximability}.
The intuition behind the notion of strict implementation is that
it allows one to modify instances while keeping control over costs
of solutions. For example, assume that we have a constraint
$g(u,v)$ in an instance $\I$ of {\sc Max CSP}, and there is a
strict 2-implementation
$g(y_1,y_2)+1=\max_{z}{(g_1(y_1,z)+g_2(z,y_2))}$. Then the
constraint $g(u,v)$ can be replaced by two constraints $g_1(u,z)$,
$g_2(z,v)$ such that $z$ does not appear in $\I$, and we know that
every solution of cost $c$ to $\I$ can be modified (by choosing an
appropriate value for $z$) to a solution of cost $c+1$ to the new
instance.

We say that a collection of predicates ${\cal F}$ {\em strictly
implements} a predicate $g$ if,  for some $\alpha\in\Z$, there
exists a strict $\alpha$-implementation of $g$ using predicates
only from ${\cal F}$. In this case we write $\F
\stackrel{s\phantom{8pt}}{\Longrightarrow_\alpha} f$. We write $\F
\stackrel{s\phantom{8pt}}{\Longrightarrow} f$ if $\F
\stackrel{s\phantom{8pt}}{\Longrightarrow_\alpha} f$ for some
$\alpha$. It is not difficult to show that if $f$ can be obtained
from $\F$ by a series of strict implementations then it can also
be obtained by a single strict implementation (for the Boolean
case, this is shown in Lemma~5.8~\cite{Creignouetal:siam01}). In
this paper, we will use about 60 specific strict implementations
for the case when $|D|=4$. Each of them can be straightforwardly
verified by hand, or by a simple computer program\footnote{An
example of such a program can be obtained from the authors or be
anonymously downloaded from
\texttt{http://www.ida.liu.se/\~{}petej/supermodular.html}.}.

The following lemma is a simple (but important) example of how
strict implementations work.

\begin{lemma}\label{unfromconst}
$\C_D$ strictly implements every predicate in $\U_D$.
\end{lemma}

\begin{proof}
It is easy to see that, for any $D'\sse D$, $u_{D'}(x)=\sum_{d\in
D'}{u_{d}(x)}$ is a strict 1-implementation.
\end{proof}

In our proofs, we will use problems with the bounded occurrence
property, so we now introduce notation for such problems.

\begin{definition}[bounded occurrence problems]\label{boundedprob}
 $\MCSP{\F}-k$ will denote the problem $\MCSP\F$ restricted to
instances with the number of occurrences of variables is bounded
by $k$. We will write that $\MCSP{\F}-B$ is $\APX$-complete to
denote that $\MCSP{\F}-k$ is $\APX$-complete for some $k$.
\end{definition}

Note that, by definition, repetitions of constraints in instances
of {\sc Max CSP} are allowed. If a variable occurs $t$ times in a
constraint which appears $s$ times in an instance, then this would
contribute $t\cdot s$ to the number of occurrences of that
variable in the instance.

\begin{lemma}\label{strict}
If $\F$ strictly implements a predicate $f$, and
$\MCSP{\F\cup\{f\}}-B$ is $\APX$-complete, then $\MCSP\F-B$ is
\APX-complete as well.
\end{lemma}

\begin{proof}
This lemma for the Boolean case, but without the assumption on
bounded occurrences, is Lemma~5.18 in~\cite{Creignouetal:siam01}.
Our proof is almost identical to the proof of Lemma~5.18
in~\cite{Creignouetal:siam01}, and it uses the same
$AP$-reduction. Essentially, we only need to verify that the
mapping $F$ in this reduction preserves the bounded occurrence
property.

Let $k$ be a number such that $\MCSP{\F\cup\{f\}}-k$ is
\APX-complete and let $\alpha\in\Z$ be such that $\F
\stackrel{s\phantom{8pt}}{\Longrightarrow_\alpha} f$. Take an
arbitrary instance $\I$ of $\MCSP{\F\cup\{f\}}-k$. Note that every
predicate in $\F$ can be (trivially) strictly $\alpha$-implemented
from $\F$ in such a way that each auxiliary variable appears only
once in the strict implementation (simply use any satisfiable
collection of $\alpha-1$ constraints with no repetitions of
variables); this is a small technicality which ensures uniformity in
the following transformation of instances. Replace every constraint
in $\I$
by a set of constraints appearing in the right-hand side of its
strict $\alpha$-implementation from $\F$, keeping the same primary
variables and using fresh copies of auxiliary variables every
time. Denote the obtained instance by $\I'$. The function $F$ in
this $AP$-reduction will be such that $F(\I)=\I'$ for all $\I$.
Let $t$ be the maximum number of occurrences of a variable
(primary or auxiliary) in the right-hand side of the strict
implementation of $f$ from $\F$. It is clear that $\I'$ is an
instance of $\MCSP\F$, and that the number of occurrences of any
variable in $\I'$ is bounded by $k'=tk$.

Let $V'$ be the set of variables in $\I'$. Let $\phi':V'\rightarrow
D$ be an $r$-approximate solution to $\I'$. The mapping $G$ uses two
possible solutions to $\I$ and takes the better of the two. The
first solution is $\phi'|_V$, while the second is a solution
satisfying $\beta=\frac{q}{|D|^a}$ constraints which exists by
Lemma~\ref{genapprox} (here $a$ is the maximum arity of constraints
in $\F\cup\{f\}$).

One can show, by literally repeating the argument in the proof of
Lemma~5.18 in~\cite{Creignouetal:siam01}, that $G(\phi')$ is an
$r'$-approximate solution to $\I$ where $r'\le 1+\gamma(r-1)$ with
$\gamma=\beta(\alpha-1)+1$.

We have constructed an $AP$-reduction from $\MCSP{\F\cup\{f\}}-k$
to $\MCSP\F-k'$, thus proving the lemma.
\end{proof}

Lemma~\ref{strict} will be used as follows in our
\APX-completeness proofs: if $\F'$ is a fixed finite collection of
predicates each of which can be strictly implemented by $\F$ then
we can assume that $\F'\sse \F$. For example, if $\F$ contains a
binary predicate $f$ then we can assume, at any time when it is
convenient, that $\F$  also contains $f'(x,y)=f(y,x)$, since this
equality is a strict 1-implementation of $f'$.

\medskip

Finally, we will use a technique based on domain restriction. For
a subset $D'\sse D$, let $\F|_{D'}=\{f|_{D'}\mid f\in \F \mbox{\
and } f|_{D'} \mbox{ is non-trivial}\}$.

\begin{lemma}\label{domrest}
Let $D'\subseteq D$ and $u_{D'}\in \F$. If $\MCSP{\F|_{D'}}-B$ is
\APX-complete then so is $\MCSP\F-B$.
\end{lemma}

\begin{proof}
Let $k$ be a bound on the number of occurences such that
$\MCSP{\F|_{D'}}-k$ is \APX-complete. We establish an
$L$-reduction from $\MCSP{\F|_{D'}}-k$ to $\MCSP\F-k'$ where
$k'=2k$.

An instance $\I$ of $\MCSP{\F|_{D'}}-k$ corresponding to
$f(x_1,\ldots,x_n)=\sum_{i=1}^q f_i({\bf x}_i)$  will be mapped to
an instance $\I'$ corresponding to
$f'(x_1,\ldots,x_n)=\sum_{i=1}^q f'_i({\bf
x}_i)+k\sum_{i=1}^{n}{u_{D'}(x_i)}$ where each $f'_i\in \F$ is
such that $f'_i|_{D'}=f_i$. We may without loss of generality
assume that all $n$ variables $x_i$ actually appear in constraint
scopes in $\I$. Note that $\I'$ is indeed an instance of
$\MCSP\F-k'$.

Let $V=\{x_1\zd x_n\}$ and fix an element $d\in D'$. If
$\phi':V\rightarrow D$ is a solution to $\I'$, then it is modified
to a solution to $\I$ as follows: set $\phi(x_i)=d$ whenever
$\phi'(x_i)\not\in D'$, and $\phi(x_i)=\phi'(x_i)$ otherwise.

We will show that this pair of mappings is an $L$-reduction for
suitable $\alpha$ and $\beta$.

Note that, for any solution to $\I'$, changing all values outside
of $D'$ to any values in $D'$ can only increase the cost of the
solution. This follows from the fact that, by changing any value
outside of $D'$ to a value in $D'$, we can lose at most $k$
satisfied constraints, but we satisfy $k$ constraints of the form
$u_{D'}(x)$. It follows that $Opt(\I')=Opt(\I)+kn$.

\smallskip

Let $a$ be the maximum arity of constraints in $\F|_{D'}$.  Let
$c=\frac{1}{|D|^a}$. Then we have $c\cdot q\le Opt(\I)$ by
Lemma~\ref{genapprox}. Set $\alpha=\frac{ak}{c}+1$. Note that we
have $n\le aq$ because the total length of constraint scopes in
$\I$ is at least $n$ and at most $aq$. Since $n\le aq\le
\frac{aOpt(\I)}{c}$, we have
\[ Opt(\I')=Opt(\I)+kn\le Opt(\I)+k\frac{aOpt(\I)}{c}=\alpha\cdot
Opt(\I),\]
so the first property of an $L$-reduction is satisfied.

We will now show that the second property is satisfied with
$\beta=1$. Let $\phi'$ and $\phi$ be solutions to $\I'$ and $\I$,
respectively, such as described above.

Let $V_1$ be the set of variables which $\phi'$ sends to
$D\setminus D'$, and $V_2$ the variables sent to $D'$; set
$r=|V_2|$. Divide all constraints in $\I'$ into three pairwise
disjoint groups: $C_1$ consists of all constraints $f_i({\bf
x}_i)$ that contain at least one variable from $V_1$, $C_2$ of all
constraints $f_i({\bf x}_i)$ that use variables only from $V_2$,
and $C_3$ contains the $kn$ constraints of the form $u_{D'}(x_i)$.
Let $q_1=|C_1|$ and $q_2=|C_2|$. Furthermore, let $s_1$ and $s_2$
be the numbers of constraints in $C_1$ and $C_2$, respectively,
that are satisfied by $\phi$. By the bounded occurrence property,
we have $s_1\le q_1\le (n-r)k$. In particular, it follows that
$s_1-nk+rk\le 0$. Note also that $cost(\phi')=s_1+s_2+rk$ and
$s_2\le cost(\phi)$. Finally, we have
\[ Opt(\I)- cost(\phi)\le Opt(\I)-s_2= \]
\[ [Opt(\I)+nk] - [s_1+s_2+rk] + [s_1-nk+rk]\le Opt(\I')-cost(\phi').\]
\end{proof}

To make use of results
in~\cite{Cohen05:supermodular,Jonsson04:maxcsp3}, we need to
introduce some more notation.

\begin{definition}[endomorphism, core]
An {\em endomorphism} of $\F$ is a unary operation $\mu$ on $D$
such that, for all $f\in \F$  and all $(a_1\zd a_m) \in D^m$, we
have

\[f(a_1\zd a_m)=1 \Rightarrow f(\mu(a_1)\zd \mu(a_m))=1.\]

We
will say that $\F$ is a {\em core} if every endomorphism of $\F$
is injective (i.e., a permutation).

If $\mu$ is an endomorphism of $\F$ with a minimal image
$im(\mu)=D'$ then a core of $\F$, denoted $core(\F)$, is the set
$\F|_{D'}$.
\end{definition}
The intuition here is that if $\F$ is not a core then
it has a non-injective endomorphism $\mu$, which implies that, for
every assignment $\phi$, there is another assignment $\mu\phi$
that satisfies all constraints satisfied by $\phi$ and uses only a
restricted set of values, so the problem is equivalent to a
problem over this smaller set. As in the case of graphs, all cores
of $\F$ are isomorphic, so one can speak about {\em the} core of
$\F$. The following rather simple corollary from
Lemma~\ref{domrest} will be useful in our proofs.

\begin{corollary}\label{core}
Assume that $\F'=core(\F)$ and $\MCSP{\F'}-B$ is \APX-complete.
Then $\MCSP\F-B$ is \APX-complete as well.
\end{corollary}

\begin{exmp}
Every set $\F$ containing $\C_D$ is a core because the only
endomorphism of $\C_D$ is the identity operation.
\end{exmp}

\section{Supermodularity and Monge properties} \label{supmoddefsection}

\subsection{Basics of supermodularity}

In this section we discuss the well-known combinatorial algebraic
property of supermodularity~\cite{Topkis98:book} which will play a
crucial role in classifying the approximability of {\sc Max CSP}
problems.

A partial order on a set $D$ is called a {\em lattice order} if,
for every $x,y \in D$, there exists a greatest lower bound $x
\sqcap y$ and a least upper bound $x \sqcup y$. The corresponding
algebra ${\cal L}=(D,\sqcap,\sqcup)$ is called a {\em lattice}.
For tuples ${\bf a}=(a_1,\ldots,a_n)$, ${\bf b}=(b_1,\ldots,b_n)$
in $D^n$, let ${\bf a} \sqcap {\bf b}$ and ${\bf a} \sqcup {\bf
b}$ denote the tuples $(a_1 \sqcap b_1,\ldots,a_n \sqcap b_n)$ and
$(a_1 \sqcup b_1,\ldots,a_n \sqcup b_n)$, respectively.

\begin{definition}[supermodular function]
Let ${\cal L}$ be a lattice on $D$. A function $f:D^n \rightarrow
\Z$ is called {\em supermodular} on ${\cal L}$ if

\[f({\bf a})+f({\bf b}) \leq
f({\bf a} \sqcap {\bf b})+f({\bf a} \sqcup {\bf b}) \; \;
\mbox{for all ${\bf a},{\bf b} \in D^n$}.\]
%
\end{definition}

Note that predicates are functions, so it makes sense to consider
supermodular predicates. We say that $\F\sse R_D$ is supermodular
on $\Lat$ if every $f\in\F$ has this property.


A finite lattice $\Lat=(D,\sqcap,\sqcup)$ is {\em distributive} if
and only if it can be represented by subsets of a set $A$, where the
operations $\sqcap$ and $\sqcup$ are interpreted as set-theoretic
intersection and union, respectively. Totally ordered lattices, or
{\em chains}, will be of special interest in this paper. Note that,
for chains, the operations $\sqcap$ and $\sqcup$ are simply $\min$
and $\max$. Hence, the supermodularity property for an $n$-ary
function $f$ on a chain is expressed as follows:
\[f(a_1\zd a_n)+f(b_1\zd b_n) \leq \]
\[f(\min(a_1,b_1)\zd \min(a_n,b_n))+f(\max(a_1,b_1)\zd
\max(a_1,b_1))\]
for all $a_1\zd a_n,b_1\zd b_n$.

\begin{exmp}\label{supmodexample}$~$

1) The disequality predicate $neq_D$ is not supermodular on any
chain on $D$. Take two elements $d_1,d_2\in D$ such that
$d_1<d_2$. Then
$$neq_D(d_1,d_2)+neq_D(d_2,d_1)=2\not\le 0= neq_D(d_1,d_1)+neq_D(d_2,d_2).$$

2) Fix a chain on $D$ and let ${\bf a}, {\bf b}$ be arbitrary
elements of $D^2$. Consider the binary predicate $f_{{\bf a}}$,
$f^{{\bf b}}$ and $f_{{\bf a}}^{{\bf b}}$ defined by the rules
\begin{eqnarray*}
f_{{\bf a}}(x,y)=1 & \Leftrightarrow & (x,y)\le {\bf a},\\
f^{{\bf b}}(x,y)=1 & \Leftrightarrow & (x,y)\ge {\bf b},\\
f_{{\bf a}}^{{\bf b}}(x,y)=1 & \Leftrightarrow & (x,y)\le {\bf a}
\mbox{ or } (x,y)\ge {\bf b},
\end{eqnarray*}
where the order on $D^2$ is component-wise.  It is easy to check
that every predicate defined above in this part of the example is
supermodular on the chain. Note that such predicates were
considered in~\cite{Cohen05:supermodular} where they were called
generalized 2-monotone. We will see later in this subsection
(Lemma~\ref{mongestru}) that such predicates are generic
supermodular binary predicates on a chain.
\end{exmp}

We will now make some simple, but useful, observations.

\begin{observation}\label{spmodobs}
$~$

\begin{enumerate}
\item Any chain is a distributive lattice.


\item Any unary predicate on $D$ is supermodular on any chain on
$D$.

\item  A predicate is supermodular on a chain if and only if it is
supermodular on its dual chain (obtained by reversing the order).
\end{enumerate}
\end{observation}

Given a chain in $D$, any binary function $f$ on $D$ can be
represented as a $|D| \times |D|$ matrix $M$ such that
$M(x,y)=f(x,y)$; here the chain indicates the order of indices of
$M$, and $M(x,y)$ is the entry in row $x$ and column $y$ of $M$.
Note that this matrix is essentially the table of values of the
predicate. For example, some binary predicates on $D=\{0,1,2,3\}$
that are supermodular on the chain $0<1<2<3$ are listed in
Fig.~\ref{badpredf1} (these predicates will be used later in the
proof of Theorem~\ref{main}). Note that all predicates in
Fig.~\ref{badpredf1} have the form described in
Example~\ref{supmodexample}(2). For example, $h_2$ is
$f_{(0,1)}^{(3,3)}$ and $h_{17}$ is $f_{(2,1)}^{(1,3)}$.

\begin{figure}[h]
\setlength{\unitlength}{4.42mm}
\begin{center}
\begin{picture}(2.5,3)(0,0)
\put(0,0.85){\scriptsize $h_{1}$} \put(1.1,1.6){\scriptsize 1000}
\put(1.1,1.1){\scriptsize 0000} \put(1.1,0.6){\scriptsize 0000}
\put(1.1,0.1){\scriptsize 0001}
\end{picture}
\begin{picture}(2.5,3)(0,0)
\put(0,0.85){\scriptsize $h_{2}$} \put(1.1,1.6){\scriptsize 1100}
\put(1.1,1.1){\scriptsize 0000} \put(1.1,0.6){\scriptsize 0000}
\put(1.1,0.1){\scriptsize 0001}
\end{picture}
\begin{picture}(2.5,3)(0,0)
\put(0,0.85){\scriptsize $h_{3}$} \put(1.1,1.6){\scriptsize 1110}
\put(1.1,1.1){\scriptsize 0000} \put(1.1,0.6){\scriptsize 0000}
\put(1.1,0.1){\scriptsize 0001}
\end{picture}
\begin{picture}(2.5,3)(0,0)
\put(0,0.85){\scriptsize $h_{4}$} \put(1.1,1.6){\scriptsize 1100}
\put(1.1,1.1){\scriptsize 1100} \put(1.1,0.6){\scriptsize 0000}
\put(1.1,0.1){\scriptsize 0001}
\end{picture}
\begin{picture}(2.5,3)(0,0)
\put(0,0.85){\scriptsize $h_{5}$} \put(1.1,1.6){\scriptsize 1110}
\put(1.1,1.1){\scriptsize 1110} \put(1.1,0.6){\scriptsize 0000}
\put(1.1,0.1){\scriptsize 0001}
\end{picture}
\begin{picture}(2.5,3)(0,0)
\put(0,0.85){\scriptsize $h_{6}$} \put(1.1,1.6){\scriptsize 1110}
\put(1.1,1.1){\scriptsize 1110} \put(1.1,0.6){\scriptsize 1110}
\put(1.1,0.1){\scriptsize 0001}
\end{picture}
\begin{picture}(2.5,3)(0,0)
\put(0,0.85){\scriptsize $h_{7}$} \put(1.1,1.6){\scriptsize 1100}
\put(1.1,1.1){\scriptsize 0000} \put(1.1,0.6){\scriptsize 0001}
\put(1.1,0.1){\scriptsize 0001}
\end{picture}
\begin{picture}(2.5,3)(0,0)
\put(0,0.85){\scriptsize $h_{8}$} \put(1.1,1.6){\scriptsize 1110}
\put(1.1,1.1){\scriptsize 0000} \put(1.1,0.6){\scriptsize 0001}
\put(1.1,0.1){\scriptsize 0001}
\end{picture}
\begin{picture}(2.5,3)(0,0)
\put(0,0.85){\scriptsize $h_{9}$} \put(1.1,1.6){\scriptsize 1000}
\put(1.1,1.1){\scriptsize 1000} \put(1.1,0.6){\scriptsize 0001}
\put(1.1,0.1){\scriptsize 0001}
\end{picture}
\begin{picture}(2.5,3)(0,0)
\put(0,0.85){\scriptsize $h_{10}$} \put(1.1,1.6){\scriptsize 1100}
\put(1.1,1.1){\scriptsize 1100} \put(1.1,0.6){\scriptsize 0001}
\put(1.1,0.1){\scriptsize 0001}
\end{picture}
\begin{picture}(2.5,3)(0,0)
\put(0,0.85){\scriptsize $h_{11}$} \put(1.1,1.6){\scriptsize 1110}
\put(1.1,1.1){\scriptsize 1110} \put(1.1,0.6){\scriptsize 0001}
\put(1.1,0.1){\scriptsize 0001}
\end{picture}
\begin{picture}(2.5,3)(0,0)
\put(0,0.85){\scriptsize $h_{12}$} \put(1.1,1.6){\scriptsize 1110}
\put(1.1,1.1){\scriptsize 0001} \put(1.1,0.6){\scriptsize 0001}
\put(1.1,0.1){\scriptsize 0001}
\end{picture}
\begin{picture}(2.5,3)(0,0)
\put(0,0.85){\scriptsize $h_{13}$} \put(1.1,1.6){\scriptsize 1000}
\put(1.1,1.1){\scriptsize 1001} \put(1.1,0.6){\scriptsize 0001}
\put(1.1,0.1){\scriptsize 0001}
\end{picture}
\begin{picture}(2.5,3)(0,0)
\put(0,0.85){\scriptsize $h_{14}$} \put(1.1,1.6){\scriptsize 1100}
\put(1.1,1.1){\scriptsize 1101} \put(1.1,0.6){\scriptsize 0001}
\put(1.1,0.1){\scriptsize 0001}
\end{picture}
\begin{picture}(2.5,3)(0,0)
\put(0,0.85){\scriptsize $h_{15}$} \put(1.1,1.6){\scriptsize 1000}
\put(1.1,1.1){\scriptsize 1001} \put(1.1,0.6){\scriptsize 1001}
\put(1.1,0.1){\scriptsize 0001}
\end{picture}
\begin{picture}(2.5,3)(0,0)
\put(0,0.85){\scriptsize $h_{16}$} \put(1.1,1.6){\scriptsize 1100}
\put(1.1,1.1){\scriptsize 1100} \put(1.1,0.6){\scriptsize 1101}
\put(1.1,0.1){\scriptsize 0001}
\end{picture}
\begin{picture}(2.5,3)(0,0)
\put(0,0.85){\scriptsize $h_{17}$} \put(1.1,1.6){\scriptsize 1100}
\put(1.1,1.1){\scriptsize 1101} \put(1.1,0.6){\scriptsize 1101}
\put(1.1,0.1){\scriptsize 0001}
\end{picture}
\begin{picture}(2.5,3)(0,0)
\put(0,0.85){\scriptsize $h_{18}$} \put(1.1,1.6){\scriptsize 1100}
\put(1.1,1.1){\scriptsize 1100} \put(1.1,0.6){\scriptsize 0011}
\put(1.1,0.1){\scriptsize 0011}
\end{picture}
\end{center}
\caption{A list of predicates on $\{0,1,2,3\}$ which are
supermodular on the chain $0<1<2<3$. The predicates are
represented by tables of values.} \label{badpredf1}
\end{figure}

A square matrix $M$ is called {\em anti-Monge} (or a-Monge, for
short)\footnote{Other names used for such matrices are {\em inverse
Monge} and {\em dual Monge}.} if $M(i,s)+M(r,j)\leq M(i,j)+M(r,s)$
for all $i<r$ and $j<s$. It is well known (and easy to check) that
matrices corresponding to binary supermodular functions on a chain
are precisely the a-Monge matrices (see, e.g., Observation~6.1
in~\cite{Burkard96:Monge}). Hence, one can view the tables in
Fig.~\ref{badpredf1} as a-Monge matrices. We will be particularly
interested in binary supermodular {\em predicates} on chains, and
the next result describes the structure of 0-1 a-Monge square
matrices.

In order to make the correspondence between matrices and binary
functions more transparent, we will use the set $J=\{0\zd n-1\}$ to
number rows and columns of an $n\times n$ matrix. Let $L_n^{pq}$
denote the square 0-1 matrix of size $n$ such that $L_n^{pq}(i,j)=1$
if and only if $i\le p$ and $j\le q$. Similarly, $R_n^{st}$ denotes
the square 0-1 matrix of size $n$ such that $R_n^{st}(i,j)=1$ if and
only if $i\ge s$ and $j\ge t$.
Let $U$ and $W$ be two subsets of $J$. We denote by $M[U,W]$ the
$|U|\times |W|$ submatrix of $M$ that is obtained by deleting all
rows not contained in $U$ and all columns not in $W$. Expression
$M[U,W]=a$ will mean that all elements in the submatrix are equal to
$a$.



\begin{lemma}\label{mongestru}
A non-zero 0-1 matrix $M$ of size $n\times n$ without all-ones
rows and columns is an a-Monge matrix if and only if one of the
following holds
\begin{itemize}
\item $M=L^{pq}_n$, for some $0\leq p,q \leq n-2$, or \item
$M=R^{st}$, for some  $1\leq s,t \leq n-1$, or \item
$M=L^{pq}_n+R^{st}_n$ for some $0\leq p,q \leq n-2$ and
 $1\leq s,t \leq n-1$, with $p<s$, or $q<t$, or both.
\end{itemize}
\end{lemma}

\begin{proof}
\noindent It is easy to see that matrices $L^{pq}_n$, $R^{st}_n$
and $L^{pq}_n+R^{st}_n$ are a-Monge matrices. Assume now that
matrix $M$ is an a-Monge matrix. We consider two cases: $M(0,0)=0$
and $M(0,0)=1$.

\smallskip
\noindent In what follows we use the following {\em $0$-$1$
property} of 0-1 a-Monge matrices:
\begin{itemize}
\item if $M(i,k)=0, M(i,j)=1$ for $k<j$, then $M(l,k)=0, M(l,j)=1$
for all $l>i$;

\item if $M(i,k)=0, M(l,k)=1$ for $i<l$, then $M(i,j)=0, M(l,j)=1$
for all $j>k$;

\item if $M(l,k)=1, M(l,j)=0$ for $k<j$, then $M(i,k)=1, M(i,j)=0$
for all $i<l$;

\item if $M(i,j)=1, M(l,j)=0$ for $i<l$, then $M(i,k)=1, M(l,k)=0$
for all $k<j$.
\end{itemize}

 Let $M(0,0)=0$. It follows from the $0$-$1$ property
that then row 0 and column 0 in the matrix contain only zeros
because, otherwise, the matrix would have an all-ones row or
column.

Let $M(u,v)=1$ for some $u,v>0$. Since the row 0 and column 0
contain only zeros, the $0$-$1$ property
 yields $M(x,y)=1$ for all $x\ge u,y\ge v$ (and, in particular, $M(n-1,n-1)=1$).
 Let  $s$ be the smallest row containing one ($s>0$), and $t$ be
the smallest column containing one in row $s$. We claim that there
is  no column $j$, $j<t$, with $M(z,j)=1$ for some $z$. Indeed, in
this case the submatrix $M[\{s,z\},\{j,t\}]$ would not be an
a-Monge matrix. This completes the proof that, in the case when
$M(0,0)=0$, we have $M=R^{st}_n$, for some $1\leq s,t \leq n-1$.
If $M(0,0)=1$ and $M(n-1,n-1)=0$, then by symmetry, we have
$M=L^{pq}_n$, for some $0\leq p,q \leq n-2$.

Let us now consider the case with $M(0,0)=M(n-1,n-1)=1$. Recall
that, by assumption, every row and every column of $M$ contains at
least one 0. Let $p+1$ be the smallest row with $0$ in column 0,
and $q+1$ be the smallest column with $0$ in row 0. Then we have
$M[\{0,\ldots,p\},\{0,\ldots,q\}]=1$. Indeed, if there is
$M(i,j)=0$ for some $0<i\le p$ and $0<j\le q$ then we choose this
element to be as close to the left-top corner as possible, and the
submatrix $M[\{i-1,i\},\{j-1,j\}]$ is not a-Monge. Furthermore,
let $s-1$ be the largest row with $0$ in the last column, and
$t-1$ be the largest column with $0$ in the last row. As above, we
have $M[\{s,\ldots,n-1\},\{t,\ldots,n-1\}]=1$. Note that we have
$p<s$, or $q<t$, or both, since, otherwise, $M$ would contain an
all-ones row or column.

It follows from the $0$-$1$ property that $M(j,0)=0$ for all $j>p$
and $M(n-1,l)=0$ for all $l<t$. As above, we derive that
$M[\{p+1,\ldots,n-1\},\{0,\ldots,t\}]=0$. By similar arguments,
$M[\{0,\ldots,s\},\{q+1,\ldots,n-1\}]=0$. This completes the proof
of the lemma in the case when both $p<s$ and $q<t$. By symmetry,
it is now enough to consider the case when $p<s$ and $q\ge t$. It
remains to show that $M[\{p+1\zd s\},\{t+1\zd q\}]=0$. Assume that
$M(i,j)=1$ for some $p+1\le i\le s$ and $t+1\le q$.  Column $j$
contains a zero, that is, $M(k,j)=0$ for some $p+1\le k\le s$. It
is easy to check that if $i<k$ then $M[\{0,j\},\{i,k\}]$ is not
a-Monge. Similarly, if $k<i$ then $M[\{j,n-1\},\{k,i\}]$ is not
a-Monge. The lemma is proved.
\end{proof}


The family of $n$-ary supermodular functions on a chain was also
studied under the name of $n$-dimensional {\em anti-Monge
arrays}~\cite{Burkard96:Monge}. As a special case of Lemma~6.3
of~\cite{Burkard96:Monge}, we have the following result (see also
Observation~6.1 of~\cite{Burkard96:Monge}).

\begin{lemma} \label{binaryenough}
An $n$-ary, $n\ge 2$, function $f$ is supermodular on a fixed
chain if and only if the following holds: every binary function
obtained from $f$ by replacing any given $n-2$ variables by any
constants is supermodular on this chain.
\end{lemma}

\subsection{Supermodularity and {\sc Max
CSP}}\label{CSPandSpmodsec}

The property of supermodularity has been used to classify the
approximability of problems $\MCSP\F$ for small sets $D$ (though,
originally the classification for the case $|D|=2$ was
obtained and stated in~\cite{Creignou95:maximum,Creignouetal:siam01,%
Khanna01:approximability} without using this property).

\begin{theorem}[\cite{Cohen05:supermodular,Creignouetal:siam01,Jonsson04:maxcsp3}]\label{smallD}
Let $|D|\le 3$ and let $\F\sse R_D$ be a core. If $\F$ is
supermodular on some chain on $D$ then weighted $\MCSP\F$ belongs to
$\PO$. Otherwise, $\MCSP\F$ is $\APX$-complete.
\end{theorem}

\begin{remark}\label{smallred} It was shown in Lemma 5.37 of \cite{Creignouetal:siam01}
that, for $D=\{0,1\}$, $\F\sse R_{\{0,1\}}$ can strictly implement
$neq_2$ whenever $\MCSP\F$ is $\APX$-complete in the above theorem
(i.e. whenever $\F$ is a core that is not supermodular on any
chain). Moreover, it follows from (the proof of) Theorem
3~\cite{Jonsson04:maxcsp3} that if $|D|=3$ and $\F$ is
supermodular on some chain on $D$ then $\F\cup \C_D$ can strictly
express $neq_2$ or $neq_3$ by using a sequence of the following
operations:
\begin{itemize}
\item adding to $\F$ a predicate that can be strictly implemented
from $\F$

\item taking the core of a subset of $\F$ (i.e., replacing $\F$ by
a subset of $\F|_{D'}$ for some $D'\sse D$).
\end{itemize}
\end{remark}

It was shown in~\cite{Alimonti00:cubic} that {\sc Max Cut} remains
\APX-complete even when restricted to cubic graphs. Since {\sc Max
Cut} is the same problem as $\MCSP{\{neq_2\}}$ (see
Example~\ref{maxkcol}), it follows that $\MCSP{\{neq_2\}}-B$ is
\APX-complete. Moreover, since $neq_k|_{\{0,1\}}=neq_{2}$, it
follows from Lemma~\ref{domrest} that
$\MCSP{\{neq_k,u_{\{0,1\}}\}}-B$ is $\APX$-complete for any $k$.
Therefore, we obtain the following corollary by combining
Remark~\ref{smallred} with Lemmas~\ref{strict} and~\ref{domrest}.

\begin{corollary}\label{hardcol}
Let $|D|\le 3$ and $\F$ not supermodular on any chain on $D$. Then
the problem $\MCSP{\F\cup \U_D}-B$ is $\APX$-complete.
\end{corollary}

The tractability part of our classification is contained in the
following result:

\begin{theorem}[\cite{Cohen05:supermodular}]\label{tract}
If $\F$ is supermodular on some distributive lattice on $D$, then
weighted\\ $\MCSP\F$ is in $\PO$.
\end{theorem}

\section{Main result}\label{mainsec}

We will need the following two technical lemmas. They will be used
in our hardness proof to reduce the argument to the case when all
non-unary predicates are binary and their matrices do not contain
all-ones rows or columns.

\begin{lemma}\label{redtobin}
If $\F$ is not supermodular on any chain on $D$ then $\F\cup\U_D$
can strictly implement a collection $\F'$ of binary predicates
which is is not supermodular on any chain on $D$.
\end{lemma}

\begin{proof}
Let $f\in \F$ be {\em not} supermodular on some fixed chain. By
Observation~\ref{spmodobs}(2), $f$ is $n$-ary with $n\ge 2$. By
Lemma~\ref{binaryenough}, it is possible to substitute constants for
some $n-2$ variables of $f$ to obtain a binary predicate $f'$ which
is not supermodular on this chain. Assume without loss of generality
that these variables are the last $n-2$ variables, and the
corresponding constants are $d_3\zd d_n$, that is,
$f'(x,y)=f(x,y,d_3,\ldots,d_{n})$. Then the following is a strict
$(n-1)$-implementation of $f'$:
$$f'(x,y)+(n-2)=\max_{z_3\zd
z_n}[f(x,y,z_3,\ldots,z_{n})+u_{\{d_3\}}(z_3)+\ldots
+u_{\{d_n\}}(z_n)].$$
Repeating this for all chains on $D$, one can strictly implement a
collection $\F'$ of binary predicates that is not supermodular on
any chain.
\end{proof}

\begin{lemma}\label{noallone}[Lemma 3.3~\cite{Jonsson04:maxcsp3}]
Assume that $h\in R_D^{(2)}$ and there is $a\in D$ such that
$h(x,a)=1$ for all $x\in D$. Let $h'(x,y)=0$ if $y=a$ and
$h'(x,y)=h(x,y)$ if $y\ne a$. Then the following holds:
\begin{enumerate}
\item for any chain on $D$, $h$ and $h'$ are supermodular (or not
supermodular) on the chain simultaneously;

\item the problems $\MCSP{\{h\}\cup {\cal U}_D}$ and
$\MCSP{\{h'\}\cup {\cal U}_D}$ are $AP$-reducible to each other.
\end{enumerate}
\end{lemma}

The next result immediately follows from Corollary~\ref{01mongerest}
and Proposition~\ref{redto3} (see Appendix A), by using the
correspondence between 0-1 a-Monge matrices and binary supermodular
predicates.

\begin{proposition}\label{bigred} If $\F$ is a set of binary predicates that
is not supermodular on any chain on $D$ then there exist $\F'\sse
\F$ with $|\F'|\le 3$ and $D'\sse D$ with $|D'|\le 4$ such that
$\F'|_{D'}$ is not supermodular on any chain on $D'$.
\end{proposition}

Recall that all predicates from $\C_D$ are supermodular on any
chain on $D$. We will now prove our main result:

\begin{theorem}\label{main}
If $\F$ is supermodular on some chain on $D$
then weighted $\MCSP{\F\cup\C_D}$ belongs to $\PO$. Otherwise,
$\MCSP{\F\cup\C_D}-B$ is $\APX$-complete.
\end{theorem}

\begin{proof}
The tractability part of the proof follows immediately from
Theorem~\ref{tract} (see also Observation~\ref{spmodobs}(1)). By
Lemmas~\ref{unfromconst} and~\ref{strict}, it is sufficient to
prove the hardness part for sets of the form $\F\cup\U_D$. We will
show that  $\{neq_2\}$ can be obtained from $\F\cup\U_D$ by using
the following two operations:
\begin{enumerate}
\item replacing $\F\cup\U_D$ by a subset of $\F\cup \U_D\cup
\{f\}$ where $f$ is a predicate that can be strictly implemented
from $\F\cup\U_D$;

\item replacing $\F\cup\U_D$ by a subset of $\F|_{D'}\cup \U_{D'}$
for some $D'$.
\end{enumerate}
By Example~\ref{maxkcol} and Lemmas~\ref{strict}
and~\ref{domrest}, this will establish the result.

It follows from Lemmas~\ref{redtobin} and~\ref{strict} that it is
sufficient to prove the hardness part of Theorem~\ref{main}
assuming that $\F$ contains only binary predicates. Now,
Proposition~\ref{bigred} and Lemma~\ref{domrest} imply that, in
addition, we can assume that $|\F|\le 3$ and $|D|\le 4$. Note that
the case $|D|\le 3$ is already considered in
Corollary~\ref{hardcol} (see also Remark~\ref{smallred}), so it
remains to consider the case $|D|=4$; we can without loss of
generality assume in the rest of the proof that $D=\{0,1,2,3\}$.
Moreover, due to Lemma~\ref{domrest}, we may consider only sets
$\F$ satisfying the following condition:

\[\mbox{for any proper subset $D'\subset D$, $\F|_{D'}$ is supermodular on some chain on $D'$. \ \ \ \  $(\ast)$}\]

\smallskip

We can assume that $\F$ is minimal with respect to inclusion, that
is, every proper non-empty subset of $\F$ is supermodular on some
chain on $D$. We will consider three cases depending on the number
of predicates in $\F$. Note that, by Lemma~\ref{noallone}, we can
without loss of generality assume that none of the predicates in
$\F$ has a matrix containing an all-ones row or column (this
property does not depend on the order of indices in the matrix).

We prove the result by using a computer-generated case analysis in
each of the three cases. In each case, we first produce a list of
all possible sets $\F$ with the above restrictions, then optimize
the list by using various symmetries, and, finally, for each
remaining set $\F$, provide a strict implementation of a set $\F'$
that is known to have an \APX-hard $\MCSP{\F'}$ problem. To
compactly describe such symmetries, we introduce some notation.
Let $\pi$ be a permutation on $D$ and $f$ a binary predicate on
$D$. Then, we define $\pi(f)$ to be the predicate such that
$\pi(f)(a,b)=1$ if and only if $f(\pi(a),\pi(b))=1$ for all $a,b
\in D$; in this case we say that the predicate $\pi(f)$ is
isomorphic to $f$. We also define the predicate $f^t$ so that
$f^t(a,b)=1$ if and only if $f(b,a)=1$ for all $a,b \in D$ (this
corresponds to transposing the matrix of $f$). We say that a
predicate of the form $\pi(f^t)$ is anti-isomorphic to $f$.

\smallskip

\underline{Case 1.} $|\F|=1$.\\
First, we use exhaustive search to generate the list of all binary
predicates $f$ on $D$ that (a) do not have all-ones rows or
columns, (b) are not supermodular on any chain on $D$, and (c)
$\F=\{f\}$ satisfies condition $(\ast)$. Moreover, we may consider
predicates only up to isomorphism and anti-isomorphism. Thus, this
list is then processed as follows: for every predicate $f$ in the
list, in order, remove all predicates below $f$ in the list that
are isomorphic or anti-isomorphic to $f$.

Clearly, it is sufficient to prove the hardness result for all
predicates that remain in the optimized list. Since there are only
$2^{16}=65536$ predicates to check, it is clear that generating
and optimizing the list can easily be (and actually was) performed
by a computer. The optimized list contains only 27 predicates
which are given in Fig.~\ref{badpred}.

\begin{figure}[h]
\setlength{\unitlength}{4.42mm}
\begin{center}
\begin{picture}(2.5,3)(0,0)
\put(0,0.85){\scriptsize $h'_{1}$} \put(1.1,1.6){\scriptsize 1000}
\put(1.1,1.1){\scriptsize 0110} \put(1.1,0.6){\scriptsize 1000}
\put(1.1,0.1){\scriptsize 0000}
\end{picture}
\begin{picture}(2.5,3)(0,0)
\put(0,0.85){\scriptsize $h'_{2}$} \put(1.1,1.6){\scriptsize 1000}
\put(1.1,1.1){\scriptsize 1101} \put(1.1,0.6){\scriptsize 1000}
\put(1.1,0.1){\scriptsize 0000}
\end{picture}
\begin{picture}(2.5,3)(0,0)
\put(0,0.85){\scriptsize $h'_{3}$} \put(1.1,1.6){\scriptsize 1001}
\put(1.1,1.1){\scriptsize 0111} \put(1.1,0.6){\scriptsize 1110}
\put(1.1,0.1){\scriptsize 1001}
\end{picture}
\begin{picture}(2.5,3)(0,0)
\put(0,0.85){\scriptsize $h'_{4}$} \put(1.1,1.6){\scriptsize 1010}
\put(1.1,1.1){\scriptsize 0101} \put(1.1,0.6){\scriptsize 1010}
\put(1.1,0.1){\scriptsize 1000}
\end{picture}
\begin{picture}(2.5,3)(0,0)
\put(0,0.85){\scriptsize $h'_{5}$} \put(1.1,1.6){\scriptsize 1010}
\put(1.1,1.1){\scriptsize 0110} \put(1.1,0.6){\scriptsize 0000}
\put(1.1,0.1){\scriptsize 0000}
\end{picture}
\begin{picture}(2.5,3)(0,0)
\put(0,0.85){\scriptsize $h'_{6}$} \put(1.1,1.6){\scriptsize 1010}
\put(1.1,1.1){\scriptsize 0111} \put(1.1,0.6){\scriptsize 1010}
\put(1.1,0.1){\scriptsize 1000}
\end{picture}
\begin{picture}(2.5,3)(0,0)
\put(0,0.85){\scriptsize $h'_{7}$} \put(1.1,1.6){\scriptsize 1010}
\put(1.1,1.1){\scriptsize 0111} \put(1.1,0.6){\scriptsize 1110}
\put(1.1,0.1){\scriptsize 1000}
\end{picture}
\begin{picture}(2.5,3)(0,0)
\put(0,0.85){\scriptsize $h'_{8}$} \put(1.1,1.6){\scriptsize 1011}
\put(1.1,1.1){\scriptsize 0101} \put(1.1,0.6){\scriptsize 1010}
\put(1.1,0.1){\scriptsize 0000}
\end{picture}
\begin{picture}(2.5,3)(0,0)
\put(0,0.85){\scriptsize $h'_{9}$} \put(1.1,1.6){\scriptsize 1011}
\put(1.1,1.1){\scriptsize 0111} \put(1.1,0.6){\scriptsize 0010}
\put(1.1,0.1){\scriptsize 0000}
\end{picture}
\begin{picture}(2.5,3)(0,0)
\put(0,0.85){\scriptsize $h'_{10}$} \put(1.1,1.6){\scriptsize 1011}
\put(1.1,1.1){\scriptsize 0111} \put(1.1,0.6){\scriptsize 0010}
\put(1.1,0.1){\scriptsize 0001}
\end{picture}
\begin{picture}(2.5,3)(0,0)
\put(0,0.85){\scriptsize $h'_{11}$} \put(1.1,1.6){\scriptsize 1011}
\put(1.1,1.1){\scriptsize 0111} \put(1.1,0.6){\scriptsize 0011}
\put(1.1,0.1){\scriptsize 0000}
\end{picture}
\begin{picture}(2.5,3)(0,0)
\put(0,0.85){\scriptsize $h'_{12}$} \put(1.1,1.6){\scriptsize 1011}
\put(1.1,1.1){\scriptsize 0111} \put(1.1,0.6){\scriptsize 0110}
\put(1.1,0.1){\scriptsize 1001}
\end{picture}
\begin{picture}(2.5,3)(0,0)
\put(0,0.85){\scriptsize $h'_{13}$} \put(1.1,1.6){\scriptsize 1011}
\put(1.1,1.1){\scriptsize 0111} \put(1.1,0.6){\scriptsize 1010}
\put(1.1,0.1){\scriptsize 0000}
\end{picture}
\begin{picture}(2.5,3)(0,0)
\put(0,0.85){\scriptsize $h'_{14}$} \put(1.1,1.6){\scriptsize 1011}
\put(1.1,1.1){\scriptsize 0111} \put(1.1,0.6){\scriptsize 1010}
\put(1.1,0.1){\scriptsize 0001}
\end{picture}
\begin{picture}(2.5,3)(0,0)
\put(0,0.85){\scriptsize $h'_{15}$} \put(1.1,1.6){\scriptsize 1011}
\put(1.1,1.1){\scriptsize 0111} \put(1.1,0.6){\scriptsize 1110}
\put(1.1,0.1){\scriptsize 0000}
\end{picture}
\begin{picture}(2.5,3)(0,0)
\put(0,0.85){\scriptsize $h'_{16}$} \put(1.1,1.6){\scriptsize 1011}
\put(1.1,1.1){\scriptsize 0111} \put(1.1,0.6){\scriptsize 1110}
\put(1.1,0.1){\scriptsize 0001}
\end{picture}
\begin{picture}(2.5,3)(0,0)
\put(0,0.85){\scriptsize $h'_{17}$} \put(1.1,1.6){\scriptsize 1011}
\put(1.1,1.1){\scriptsize 0111} \put(1.1,0.6){\scriptsize 1110}
\put(1.1,0.1){\scriptsize 1001}
\end{picture}
\begin{picture}(2.5,3)(0,0)
\put(0,0.85){\scriptsize $h'_{18}$} \put(1.1,1.6){\scriptsize 1011}
\put(1.1,1.1){\scriptsize 0111} \put(1.1,0.6){\scriptsize 1110}
\put(1.1,0.1){\scriptsize 1101}
\end{picture}
\begin{picture}(2.5,3)(0,0)
\put(0,0.85){\scriptsize $h'_{19}$} \put(1.1,1.6){\scriptsize 1011}
\put(1.1,1.1){\scriptsize 1101} \put(1.1,0.6){\scriptsize 1010}
\put(1.1,0.1){\scriptsize 0000}
\end{picture}
\begin{picture}(2.5,3)(0,0)
\put(0,0.85){\scriptsize $h'_{20}$} \put(1.1,1.6){\scriptsize 1100}
\put(1.1,1.1){\scriptsize 1101} \put(1.1,0.6){\scriptsize 1000}
\put(1.1,0.1){\scriptsize 0000}
\end{picture}
\begin{picture}(2.5,3)(0,0)
\put(0,0.85){\scriptsize $h'_{21}$} \put(1.1,1.6){\scriptsize 1101}
\put(1.1,1.1){\scriptsize 0110} \put(1.1,0.6){\scriptsize 0110}
\put(1.1,0.1){\scriptsize 1001}
\end{picture}
\begin{picture}(2.5,3)(0,0)
\put(0,0.85){\scriptsize $h'_{22}$} \put(1.1,1.6){\scriptsize 1101}
\put(1.1,1.1){\scriptsize 1100} \put(1.1,0.6){\scriptsize 0010}
\put(1.1,0.1){\scriptsize 0000}
\end{picture}
\begin{picture}(2.5,3)(0,0)
\put(0,0.85){\scriptsize $h'_{23}$} \put(1.1,1.6){\scriptsize 1101}
\put(1.1,1.1){\scriptsize 1110} \put(1.1,0.6){\scriptsize 0000}
\put(1.1,0.1){\scriptsize 0000}
\end{picture}
\begin{picture}(2.5,3)(0,0)
\put(0,0.85){\scriptsize $h'_{24}$} \put(1.1,1.6){\scriptsize 1101}
\put(1.1,1.1){\scriptsize 1110} \put(1.1,0.6){\scriptsize 0110}
\put(1.1,0.1){\scriptsize 1001}
\end{picture}
\begin{picture}(2.5,3)(0,0)
\put(0,0.85){\scriptsize $h'_{25}$} \put(1.1,1.6){\scriptsize 1110}
\put(1.1,1.1){\scriptsize 1100} \put(1.1,0.6){\scriptsize 0000}
\put(1.1,0.1){\scriptsize 0000}
\end{picture}
\begin{picture}(2.5,3)(0,0)
\put(0,0.85){\scriptsize $h'_{26}$} \put(1.1,1.6){\scriptsize 1110}
\put(1.1,1.1){\scriptsize 1100} \put(1.1,0.6){\scriptsize 1010}
\put(1.1,0.1){\scriptsize 0000}
\end{picture}
\begin{picture}(2.5,3)(0,0)
\put(0,0.85){\scriptsize $h'_{27}$} \put(1.1,1.6){\scriptsize 1110}
\put(1.1,1.1){\scriptsize 1101} \put(1.1,0.6){\scriptsize 1010}
\put(1.1,0.1){\scriptsize 0000}
\end{picture}
\end{center}
\caption{The optimized list of 27 predicates from the proof of
Case 1. The predicates are represented by tables of values.}
\label{badpred}
\end{figure}

We show, starting from $h'_{1}$ and proceeding in order, that
$\{h'_i\}\cup {\cal U}_D$ strictly implements some binary
predicate $g$ such that either, for some $D'\subset D$, the
predicate $g|_{D'}$ is not supermodular on any chain on $D'$ or
$g$ is equal to $h'_j$ for some $j<i$ (up to isomorphism and
anti-isomorphism). These implementations can be found in Appendix
B. This, together with Remark~\ref{smallred}, implies that $neq_2$
can be obtained from $\F\cup\U_D$.

\smallskip

\underline{Case 2.} $|\F|=2$.\\
Let $\F=\{f_1,f_2\}$. As in Case 1, we use exhaustive search to
generate the list of all pairs of binary predicates on $D$ such
that (a) they do not have all-ones rows or columns, (b) each of
the two predicates is supermodular on at least one chain, but
there is no chain on which they are both supermodular, and (c)
$\F$ satisfies condition $(\ast)$. Without loss of generality, we
can assume that $f_1$ is supermodular on the chain $0<1<2<3$, that
is, the matrix of $f_1$ with this order of indices is a-Monge.
Since the matrix of $f_1$ does not have all-ones row or column,
its structure is described in Lemma~\ref{mongestru}. Similarly,
the matrix of $f_2$ is a permuted a-Monge matrix, since $\pi(f_2)$
is supermodular for some permutation $\pi$.

We can also assume that the a-Monge matrices for $f_1$ and $f_2$
(with respect to the orders on which the predicates are
supermodular) have the third form ($L_4^{pq}+R_4^{st}$) from
Lemma~\ref{mongestru}. The reason is that if, say, the matrix of
$f_1$ has the form $L_4^{pq}$ for some $0\le p,q\le 2$ then
$f'_1(x,y)+1=f_1(x,y)+ u_{\{p+1\zd 3\}}(x)+u_{\{q+1\zd 3\}}(y)$ is
a strict 2-implementation of the predicate $f'$ whose matrix is
$R_4^{(p+1)(q+1)}$. Moreover, $f_1''(x,y)=f_1(x,y)+f_1'(x,y)$ is a
strict 1-implementation of a predicate whose matrix is
$L_4^{pq}+R_4^{(p+1)(q+1)}$. Hence, we can replace $f_1$ by
$f_1''$ in this pair, and show the hardness result for
$\{f_1'',f_2\}$.

It is clear that if prove the result for {\em all} pairs
$(f_1,f_2)$ with some fixed $f_1$, then this also proves the
result for all pairs with the first component $f_1^t$, or
$\pi(f_1)$, or $\pi(f^t_1)$ where $\pi(x)=3-x$. This implies that
it is sufficient to consider only predicates from
Fig.~\ref{badpredf1} as possible candidates for $f_1$. Moreover,
it can be straightforwardly checked by using a computer that if
$f_1$ is one of the predicates $h_1,h_3,h_4,h_6,h_7,h_9,h_{10}$
from Fig.~\ref{badpredf1}, then $\F=\{f_1,f_2\}$ fails to satisfy
condition $(\ast)$. Hence, all pairs $(f_1,f_2)$, where at least
one of $f_1$ and $\pi(f_2)$ (for some permutation $\pi$) coincides
with one of 7 predicates above, will not be on the list of pairs
that we need to consider.

Obviously, if we prove the result for some pair $(f_1,f_2)$ then
this also proves the result for $(f_1,f_2^t)$. Hence, provided
$f_2\ne f_2^t$, one of these two pairs can be excluded from the
list.

Now we show that predicates $h_5,h_{11},h_{12}$, and $h_{17}$ from
Fig.~\ref{badpredf1} can also be excluded from consideration
because they can strictly implement some other predicates from
Fig.~\ref{badpredf1}.
Implementations:\\
    \\
{\small $\left\{f := {\tiny \begin{array}{l}
1100\\
1101\\
1101\\
0001\\
\end{array}
}\right\} \cup$ ${\cal U}_D
\stackrel{s\phantom{8pt}}{\Longrightarrow_6}$ ${\tiny
\begin{array}{l}
1100\\
0001\\
0001\\
0001\\
\end{array}
} =: g\qquad$  $f=h_{17}, g=\pi(h_8) \mbox{ where } \pi(x)=3-x$}
\\
{\small
$g(x,y)+5=max_{z,w}[f(z,w)+f(z,y)+f(x,z)+f(x,w)+u_{\{0,3\}}(z)+u_{\{3\}}(w)+u_{\{0\}}(x)]$}

\vskip 0.3cm

\noindent {\small $\left\{f := {\tiny \begin{array}{l}
1110\\
1110\\
0001\\
0001\\
\end{array}
}\right\} \cup$ ${\cal U}_D
\stackrel{s\phantom{8pt}}{\Longrightarrow_3}$ ${\tiny
\begin{array}{l}
1110\\
1110\\
0000\\
0001\\
\end{array}
} =: g\qquad$  $f=h_{11}, g=h_5$}
\\
{\small  $g(x,y)+2=max_z[f(z,x)+f(z,y)+f(x,z)]$}

\vskip 0.3cm

\noindent {\small $\left\{f := {\tiny \begin{array}{l}
1110\\
1110\\
0000\\
0001\\
\end{array}
}\right\} \cup$ ${\cal U}_D
\stackrel{s\phantom{8pt}}{\Longrightarrow_2}$ ${\tiny
\begin{array}{l}
1100\\
1100\\
1101\\
0001\\
\end{array}
} =: g\qquad$  $f=h_5, g=h_{16}$}
\\
{\small  $g(x,y)+1=max_z[f(x,z)+f(y,z)+u_{\{2\}}(x)]$}

\vskip 0.3cm

\noindent {\small $\left\{f := {\tiny \begin{array}{l}
1110\\
0001\\
0001\\
0001\\
\end{array}
}\right\} \cup$ ${\cal U}_D
\stackrel{s\phantom{8pt}}{\Longrightarrow_4}$ ${\tiny
\begin{array}{l}
1000\\
1001\\
1001\\
0001\\
\end{array}
} =: g\qquad$  $f=h_{12}, g=h_{15}$}
\\
{\small
$g(x,y)+3=max_z[f(z,x)+f(z,y)+f(x,z)+f(y,z)+u_{\{1\}}(z)+u_{\{1,2\}}(x)]$}

\vskip 0.3cm

\noindent As above, all pairs $(f_1,f_2)$ such that, for some
permutation $\pi$, $\pi(f_2)$ or $\pi(f_2^t)$ is one of
$h_5,h_{11},h_{12},h_{17}$, can also be excluded from the list.

Finally, we can exclude from the list all pairs isomorphic to some
pair higher up in the list. That is, we exclude pair $(f_1,f_2)$
if there is a permutation $\pi$ such that either $\pi(f_1)=f_1$
and the pair $(f_1,\pi(f_2))$ is above $(f_1,f_2)$ in the list or
if there is a permutation $\pi$ such that the pair
$(\pi(f_2),\pi(f_1))$ is above $(f_1,f_2)$ in the list (in the
latter case, $\pi(f_2)$ must be supermodular on $0<1<2<3$).


The optimized list now contains 27 pairs of predicates. In
Appendix C, we provide strict implementations for them that show
that, for each pair $(f_1,f_2)$ in this list, $\{f_1,f_2\} \cup
{\cal U}_D$ implements either a pair above it in the list or else
a binary predicate $g$ such that, for some $D'\subset D$, the
predicate $g|_{D'}$ is not supermodular on any chain on $D'$. As
in Case 1, it follows that $neq_2$ can be obtained from
$\F\cup\U_D$.

\smallskip

\underline{Case 3.} $|\F|=3$.\\
It can be checked by computer-assisted exhaustive search that
there does not exist such a set $\F$.\footnote{The authors also
have a (rather lengthy) combinatorial proof of this fact.} Simply
loop through all triples of (not necessarily distinct) binary
predicates on $\{0,1,2\}$ which are supermodular on the chain
$0<1<2$ and check that each possible extension to a triple of
pairwise distinct predicates on $D$ results in a set $\F$
satisfying one of the following conditions:

\begin{enumerate}
\item $\F$ is supermodular on some chain on $D$,

\item for some $D'\subset D$, $\F|_{D'}$ is not supermodular on
any chain on $D'$,

\item some proper subset of $\F$ is not supermodular on any chain
on $D$.
\end{enumerate}
\end{proof}

\begin{remark} \label{testsupermodularity}
Note that, for any fixed $D$, it can be checked in polynomial time
whether a given $\F$ is supermodular on some chain on $D$. That
is, given $\F$, we can check in polynomial time whether
$\MCSP{\F\cup \C_D}$ is tractable or $\APX$-complete.
\end{remark}

\section{Application to {\sc List $H$-coloring}
optimization}\label{hcolsec}

Recall that a {\em homomorphism} from a digraph $G=(V_G,A_G)$ to a
digraph $H=(V_H,A_H)$ is a mapping \mbox{$\phi:V_G\rightarrow
V_H$} such that $(\phi(v),\phi(w))\in A_H$ whenever $(v,w)\in
A_G$. In this case, the digraph $G$ is said to be {\em
$H$-colorable}. The {\sc Graph $H$-colorability} problem is, given
a digraph $G$, to decide whether it is $H$-colorable. This problem
attracts much attention in graph theory~\cite{Hell04:book}.

In this section, we consider the case when $\F$ consists of a
single binary predicate $h$. This predicate specifies a digraph
$H$ such that $V_H=D$ and $(u,v)$ is an arc in $H$ if and only if
$h(u,v)=1$. Any instance $\I=(V,C)$ of $\CSP{\{h\}}$ can be
associated with a digraph $G_\I$ whose nodes are the variables in
$V$ and whose arcs are the scopes of constraints in $C$. It is not
difficult to see that the question whether all constraints in $\I$
are simultaneously satisfiable is equivalent to the question
whether $G_\I$ is $H$-colorable. Therefore, the problem
$\CSP{\{h\}}$ is precisely the {\sc Graph $H$-colorability}
problem for the digraph $H$. The problems $\CSP{\{h\}\cup \U_D}$
and $\CSP{\{h\}\cup \C_D}$ are equivalent to the {\sc List
$H$-coloring} and {\sc $H$-retraction} problems, respectively. In
the former problem, every vertex of an input digraph $G$ gets a
list of allowed target vertices in $H$, and the question is
whether $G$ has an $H$-coloring subject to the list constraints.
The latter problem is the same except that each list contains
either one or all vertices of $H$. These problems also attract
much attention in graph theory~\cite{Hell04:book}.

The problem $\MCSP{\{h\}\cup \U_D}$ can then be viewed as the {\sc
List $H$-coloring} optimization problem: for every vertex $v$ of
an input digraph $G$, there is a list $L_v\sse V_H$ along with a
function $\rho_v:L_v\rightarrow \Z$ that indicates the `score'
which a mapping $V_G\rightarrow V_H$ gets if it sends $v$ to a
certain vertex (if a mapping sends $v$ to a vertex outside of
$L_v$ then this adds nothing to the `cost' of this mapping). Then
the goal is to maximize the combined `cost' of such a mapping
which is obtained by adding weights of preserved arcs and `scores'
from the lists. The `score' functions $\rho_v$ arise as the result
of the possible presence in $C$ of several weighted constraints of
the form $u_{D'}(v)$ for different $D'\subseteq D$ and the same
$v$. Thus, Theorem~\ref{main} in the case when $\F=\{h\}$ presents
a complexity classification of list $H$-coloring optimization
problems. Digraphs $H$ corresponding to the tractable cases of
this problem are the digraphs that have an a-Monge adjacency
matrix under some total ordering on $V_H$ (note that this property
of digraphs can be recognised in polynomial time, e.g., by using
Proposition~\ref{bigred}). Such matrices without all-one rows or
columns are described in Lemma~\ref{mongestru}. It remains to note
that, as is easy to see, replacing either some all-zero row or
some all-zero columns with all-one ones does not affect the
property of being a-Monge.

We remark that another problem related to optimizing list
homomorphisms between graphs was recently considered
in~\cite{Gutin05:lora}, in connection with some problems arising
in defence logistics.
%

\section*{Acknowledgements}

The authors are thankful to Gerhard Woeginger for encouraging this
collaboration and to Johan H\aa stad for suggesting to use the
bounded occurrence property in our proofs.


\newpage

\appendix
\section{Appendix A: Permuted a-Monge matrices}

In this appendix, we prove results about a-Monge matrices that
will imply, via the correspondence between binary supermodular
predicates on chains and a-Monge matrices,
Proposition~\ref{bigred}:

\medskip

\noindent
If $\F$ is a set of binary predicates that is not
supermodular on any chain on $D$, then there exists $\F' \subseteq \F$
with $|\F'| \leq 3$ and $D' \subseteq D$ with $|D'| \leq 4$ such that
$\F'|_{D'}$ is not supermodular on any chain on $D'$.

\medskip

\noindent We prove this in two steps: the existence of $D'$ is
established in Section~\ref{redD} and the existence of $\F'$ in
Section~\ref{redF}. Our results about matrices will be more general
than required to prove Proposition~\ref{bigred} because will
consider general (i.e., not necessarily 0-1) matrices.

First, we need to introduce some concepts and notation. Let $M$ be
an $n \times n$ matrix. If there is a permutation $\pi$ that {\em
simultaneously} permutes rows and columns of $M$ so that the
resulting matrix is an a-Monge matrix, then the matrix $M$ is
called a {\em permuted} a-Monge matrix and the permutation is
called an {\em a-Monge permutation} for $M$. Note that we will
often use the term `permutation' as a synonym for `linear
(re-)ordering'. Given a set of indices $I=\{i_1,\ldots,i_k\}
\subseteq J=\{0\zd n-1\}$, we use notation $M[I]$ for the
sub-matrix $M[I,I]$.  We say that $M[I]$ is permuted according to
a permutation $\seq{s_1,\ldots,s_k}$, where
$I=\{s_1,\ldots,s_k\}$, if row (column) $s_1$ is the first row
(column) in the permuted matrix, $s_2$ is the second row (column),
and so on. If $n\leq 4$ and $M$ is not a permuted a-Monge matrix,
then $M$ is called a {\em bad} matrix.

A row $i$ {\em precedes\/} a row $j$ in $M$ ($i\prec j$ for
short), if row $i$ occurs before row $j$ in $M$. If $i$ precedes
$j$ in a permutation $\pi$, then  we write $i\prec_{\pi} j$. When
the permutation $\pi$ is understood from the context, we simply
write $i \prec j$. If $\pi$ is an a-Monge permutation for the
matrix $M$, then the {\em reverse} of $\pi$, $\pi^-$ defined as
$\pi^-(i)=\pi(n-1-i)$, is also an a-Monge permutation. Therefore,
given two indices $i$ and $j$, we can always assume that $i$
precedes $j$ in an a-Monge permutation (if there is any).

Denote by $\Delta(i,j,k,l)$, for $ i,j,k,l \in J $, an algebraic
sum that involves four entries of the matrix $M$:
$\Delta(i,j,k,l)=M(i,k)+M(j,l)-M(i,l)-M(j,k)$. Given a permutation
$\pi$ for permuting rows and columns in $M$, we use a similar
notation for the sums in the permuted matrix: $
\Delta(i,j,k,l,\pi)=M(\pi(i),\pi(k))+M(\pi(j),\pi(l))-M(\pi(i),\pi(l))
-M(\pi(j),\pi(k))$.
 For $k=i$ and $l=j$, we use
simplified notation
 $\Delta(i,j)=\Delta(i,j,i,j)$, for $i,j \in J$.
Matrix $M$ is an a-Monge matrix if and only if $\Delta(i,j,k,l)\ge
0$ for all $i<j$ and $k<l$, and $M$ is permuted a-Monge if and
only if there exists a permutation $\pi$ such that
$\Delta(i,j,k,l,\pi)\ge 0$ for all $i<j$ and $k<l$. It is easy to
check that
\begin{equation}\label{eq:delta}
 \Delta(i,j,k,l)=\sum_{s=i,\ldots,j-1;
t=k,\ldots,l-1}\Delta(s,s+1,t,t+1).
\end{equation}
Therefore, given a permutation $\pi$ and matrix $M$, it can be
checked in $O(n^2)$ time whether $\pi$ is an a-Monge permutation
for the matrix $M$.

We will say that row (column) $s$ is equivalent to row
(respectively, column) $t$, if $\Delta(s,t,k,l)=0$ (respectively,
$\Delta(k,l,s,t)=0$) for all $k,l$. It can easily be shown that if
rows $s$ and $t$ are equivalent, then $M(s,i)=M(t,i)+\alpha_{st}$,
for all $i$ and some constant $\alpha_{st}$.
 Hence, after
subtracting $\alpha_{st}$ from all elements in the row $s$
($M'(s,i)=M(s,i)-\alpha_{st}$), one gets two identical rows $s$
and $t$ ($M'(s,i)=M'(t,i)$ for all $i$). A matrix with all rows
(and all columns) equivalent is called a sum matrix: It can be
shown that in this case $M(s,t)=u_s+v_t$, for some real vectors
$u$ and $v$. Clearly, any sum matrix is a-Monge.


\subsection{Reducing the size of matrices} \label{redD}

We will first show that whenever an $n \times n$ matrix $M$ is not
a permuted a-Monge matrix, then there exists a set of indices $B$
with $|B| \leq 4$ such that $M[B]$ is a bad matrix. Our approach
to the recognition of bad matrices is loosely based on the COM
(Construct partial Orders and Merge them) algorithm, suggested in
\cite{Deineko94:general}. This algorithm constitutes a general
approach to deciding whether a given matrix, possibly with some
unknown elements, can be permuted to avoid a special set of $2
\times 2$ submatrices. In our case, these are submatrices
$M[\{i,j,k,l\}]$ with $\Delta(i,j,k,l)<0$.

We will use the idea of the COM algorithm, which goes as follows:
given a matrix $M$, we try to construct an a-Monge permutation for
it. We start with a pair of indices $(i,j)$ which correspond to
two non-equivalent rows or columns in the matrix. We assume
further that the index $i$ precedes index $j$ in an a-Monge
permutation $\pi$.
 The assumption $i \prec_{\pi} j$ determines
the order of some other indices. Under the assumption that
$i\prec_{\pi} j$, the strict inequality $\Delta(i,j,k,l)>0$
indicates that $k\prec_{\pi} l$, while the strict inequality
$\Delta(i,j,k,l)<0$ indicates that $l\prec_{\pi} k$. (Note that
$\Delta(i,j,k,l)=-\Delta(i,j,l,k)=-\Delta(j,i,k,l)=\Delta(j,i,l,k)$
-- this property will often be used in our proofs). The obtained
information can be conveniently represented as a directed graph
$P_M$ with nodes $J$ and directed arcs corresponding to the
identified precedence constraints together with the initial
constraint $i \prec_{\pi} j$. We then extend $P_M$ recursively to
obtain additional information about the ordering of indices.
Eventually, either $P_M$ contains an oriented cycle which signals
that the matrix is not a permuted a-Monge matrix, or we can view
$P_M$ as a partial order. This order defines a set of permutations
(i.e., linear extensions of $P_M$) which are our candidates for an
a-Monge permutation. We illustrate the COM approach with the
following example.
\smallskip

\begin{example}
Consider a submatrix $M[\{i,k,j\}]$. Schematic representation of
this sub-matrix and algebraic sums $\Delta$ is shown in
Fig.~\ref{fig:delta1}. We claim that, provided
$\Delta(i,j)=\Delta(i,k)=\Delta(k,j)= 0$, the submatrix is either
a bad matrix or a sum matrix.

It follows from
$\Delta(i,j)=\Delta(i,k)+\Delta(i,k,k,j)+\Delta(k,j,i,k)+\Delta(k,j)=
0$ that $\Delta(k,j,i,k)=-\Delta(i,k,k,j)$. Suppose that
$\Delta(k,j,i,k)\neq 0$ and $\Delta(i,k,k,j)\neq 0$. Without loss
of generality, suppose that $i\prec k$ in an a-Monge permutation
and that $\Delta(i,k,k,j)>0$. The assumption that row $i$ precedes
row $k$, together with the inequality $\Delta(i,k,k,j)>0$ yields
$k\prec j$. The assumption that column $i$ precedes column $k$
together with the inequality $\Delta(k,j,i,k)>0$ yields a
contradictory precedence $j\prec k$. Therefore $M[\{i,j,k\}]$ is a
bad matrix.

If $\Delta(k,j,i,k)= 0$ and $\Delta(i,k,k,j)= 0$, then it can
easily be shown that $M[\{i,j,k\}]$ is a sum matrix.
\end{example}

We recommend the reader to use diagrams like the one in
Fig.~\ref{fig:delta1} in the following proof, since they make
arguments more transparent.

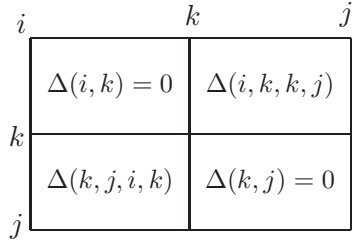
\begin{figure}
\mbox{$~$} \hspace{5.5cm} \psset{unit=0.6cm}
\begin{pspicture}(0.5,0)(3,6)
\rput[tl](0.7,5)
 {\small
  \begin{tabular}{|c|c|}
        \hline
         & \\
         $\Delta(i,k)=0$&$\Delta(i,k,k,j)$\\
    & \\
         \hline
    & \\
         $\Delta(k,j,i,k)$&$\Delta(k,j)= 0$\\
    &\\
        \hline
      \end{tabular}
  }
\rput(0.5,5.3){$i$} \rput(0.4,2.8){$k$} \rput(0.4,0.8){$j$}
\rput(4.3,5.5){$k$} \rput(7.7,5.5){$j$} 

%
%

\end{pspicture}
\caption{Schematic representation of submatrices and algebraic
sums $\Delta$.} \label{fig:delta1}
\end{figure}

\smallskip
\begin{theorem}\label{genmongerest}
If an $n\times n$ matrix $M$ is not a permuted a-Monge matrix,
then there exists a set of indices $B$ with $|B|\leq 4$, such that
$M[B]$ is a bad matrix.
\end{theorem}

\begin{proof}
We can without loss of generality assume that $M$ has no pair
$s,t$ of indices such that both rows $s,t$ are equivalent and
columns $s,t$ are equivalent. Indeed, if $s,t$ is such a pair then
it is easy to see that $M$ is permuted a-Monge if and only if
$M[J\setminus\{s\}]$ is permuted a-Monge, so we can delete row $s$
and column $s$ and continue.

First note that if there exists a pair $i,j$ such that $i\ne j$
and $\Delta(i,j)<0$, then $M[\{i,j\}]$ is a bad matrix, so we
assume further on that $\Delta(i,j)\ge 0$ for all distinct $i,j$.


Assume that $\Delta(i,j)=0$ for all $i,j$. Suppose that there
exists a triple $i,j,k$ such that $\Delta(k,j,i,k)\neq 0$ and/or
$\Delta(i,k,k,j)\neq 0$. Then, as shown in the example above,
$M[\{i,j,k\}]$ is a bad matrix in this case. Suppose instead that
$\Delta(k,j,i,k)= 0$ and $\Delta(i,k,k,j)= 0$ for all $i,k,j$. For
any $s,t,k,l$ with $s,t<k,l$, we have
$\Delta(s,t,k,l)=\Delta(s,t,t,l)-\Delta(s,t,t,k)$, and therefore
$\Delta(s,t,k,l)=0$. For any $s,t,k,l$ with $s,t>k,l$, we have
$\Delta(s,t,k,l)=\Delta(s,t,k,s)-\Delta(s,t,l,s)$, and therefore
$\Delta(s,t,k,l)=0$. It can be shown in a similar way that
$\Delta(s,t,k,l)=0$ for all $s,t,k,l$, and therefore $M$ is a sum
matrix, which is impossible because $M$ is not permuted a-Monge.

\smallskip

Assume now that $\max_{k,l}{\Delta(k,l)}>0$. We will try to
construct an a-Monge permutation for $M$, and show that such an
effort unavoidably results in the identification of a bad
submatrix in $M$. If $M$ were a permuted a-Monge matrix, then
there would exist indices $i^\star,j^\star$ and a permutation
$\pi$ with $\pi(i^\star)=0$ and $\pi(j^\star)=n-1$ such that
$\Delta(i^\star,j^\star)=\max_{k,l}{\Delta(k,l)}$ (see
equation~(\ref{eq:delta})), and also
$\Delta(i^\star,j^\star,j,j+1,\pi)\ge 0$ and
$\Delta(j,j+1,i^\star,j^\star,\pi)\ge 0$ for all
$j=0,1,\ldots,n-2$. To simplify the presentation, we assume that
$i^\star=0$ and $j^\star=n-1$ (otherwise, we renumber the rows and
columns in the matrix). The above inequalities can be rewritten as
$M(0,\pi(j))-M(n-1,\pi(j))\ge M(0,\pi(j+1))-M(n-1,\pi(j+1))$ and
$M(\pi(j),0)-M(\pi(j),n-1)\ge M(\pi(j+1),0)-M(\pi(j+1),n-1)$ for
$j=0,1,\ldots,n-2$.

So, an a-Monge permutation $\pi$ would have to sort the
differences $(M(0,i)-M(n-1,i))$ and the differences
$(M(i,0)-M(i,n-1))$, $i\neq 0,n-1$ in non-increasing order. If
there exists no permutation that sorts both sequences, then there
is a pair $i,j$ such that $M(0,i)-M(n-1,i)< M(0,j)-M(n-1,j)$
(which yields the precedence constraint $i\prec j$) and
$M(i,0)-M(i,n-1)> M(j,0)-M(j,n-1)$ (which yields the precedence
constraint $j\prec i$). This implies that matrix
$M[\{0,n-1,i,j\}]$ is a bad matrix.

Suppose now that $\Delta(0,n-1)=\max_{k,l}{\Delta(k,l)}$ and there
exists a permutation $\pi$, with $\pi(0)=0$ and $\pi(n-1)=n-1$,
that sorts both
sequences. 
Fix such a permutation and permute $M$ according to it. We can
without loss of generality assume that $M$ had this new form from
the very beginning, that is, both the sequence $(M(0,i)-M(n-1,i))$
and the sequence $(M(i,0)-M(i,n-1))$, $i\neq 0,n-1$, are already
in non-increasing order.

Since $M$ is not permuted a-Monge, we still have indices $p,q,s,t$
such that $p<q$, $s<t$, and $\Delta(p,q,s,t)<0$. It follows from
the inequality $\Delta(p,q,s,t)<0$ and from the equation
(\ref{eq:delta}) that there exists an index $i$ with $p\le i\le
q-1$, and an index $k$ with $s\le k\le t-1$, such that
$\Delta(i,i+1,k,k+1)<0$. We consider the case $i<k$: the case
$i=k$ is already eliminated and the case $i>k$ is symmetric.
%

Assume that there exist indices $i$ and $k$ with $i+1<k$ such that
$\Delta(i,i+1,k,k+1)<0$, $\Delta(i,i+1,i+1,k)>0$, and
$\Delta(i+1,k,k,k+1)>0$. We claim that $M[\{i,i+1,k,k+1\}]$ is a
bad matrix in this case. Indeed, the assumption $i\prec i+1$
 yields $i+1\prec k$ and $k+1\prec k$, and constraint
$k+1\prec k$ for the  columns $k$ and $k+1$ yields $k\prec i+1$.
This proves the claim.

Assume now that these is {\em no} pair of indices $i,k$ with
$i+1<k$ such that $\Delta(i,i+1,k,k+1)<0$,
$\Delta(i,i+1,i+1,k)>0$, and $\Delta(i+1,k,k,k+1)>0$. We claim
that then there exists a triple of indices $i,j,l$ with $i<j<l$
such that $\Delta(i,j,j,l)<0$. Indeed, we know that we have a pair
of indices $i,k$ with $i<k$ such that $\Delta(i,i+1,k,k+1)<0$. If
$i+1=k$, then our claim trivially holds with $j=k$ and $l=k+1$.
Otherwise, we have $i+1<k$ with $\Delta(i,i+1,i+1,k)\le 0$ or
$\Delta(i+1,k,k,k+1)\le 0$ (or both). If $\Delta(i,i+1,i+1,k)\le
0$ then
$\Delta(i,i+1,i+1,k+1)=\Delta(i,i+1,i+1,k)+\Delta(i,i+1,k,k+1)<0$,
so we can take $j=i+1$ and $l=k+1$ in our claim. The situation
when $\Delta(i+1,k,k,k+1)\le 0$ is treated similarly.


We consider two cases:
\begin{itemize}
\item[] \underline{Case 1} There exists a triple $i<j<l$ with
$\Delta(i,j,j,l)<0$ such that $i=0$ or $l=n-1$, or both.

We consider the case with $\Delta(0,j,j,l)<0$ (the case of
$\Delta(i,j,j,n-1)<0$ is symmetric). We claim that matrix
$M[\{0,j,l,n-1\}]$ is a bad matrix in this case. Indeed, rows and
columns in the matrix are sorted to guarantee, in particular, the
inequalities $\Delta(0,n-1,j,l)\ge 0$ and $\Delta(0,j,0,n-1)\ge
0$. It follows from the assumption $\Delta(0,j,j,l)<0$  and the
equality
 $\Delta(0,n-1,j,l)=\Delta(0,j,j,l)+\Delta(j,n-1,j,l)$ that
 $\Delta(j,n-1,j,l)>0$.
So, the assumption $0\prec j$ yields $l\prec j$, and $l\prec j$
yields $n-1\prec j$. We will show that then we have a
contradiction with the choice of $0$ and $n-1$ as a pair such that
$\Delta(0,n-1)=\max_{k,l}{\Delta(k,l)}$. If $l=n-1$ then there are
two permutations of $\{0,j,n-1\}$ compatible with the obtained
precedence constraints: $\seq{0,n-1,j}$ and $\seq{n-1,0,j}$. If
$\seq{0,n-1,j}$ is an a-Monge permutation for $M[\{0,j,n-1\}]$
then $\Delta(0,j)$ can be represented as a sum of non-negative
numbers (see equality~(\ref{eq:delta})) which include, in
particular, $\Delta(0,n-1)$ and $\Delta(n-1,j)>0$. This is a
contradiction with the choice of $0$ and $n-1$. If $\seq{n-1,0,j}$
is an a-Monge permutation for $M[\{0,j,n-1\}]$ then we get a
contradiction in a a similar way, using the fact that
$\Delta(0,j)=\Delta(0,j,0,n-1)-\Delta(0,j,j,n-1)>0$.

Assume further on that $l\ne n-1$. Since $\Delta(0,j,j,l)<0$ and
$\Delta(0,j,0,n-1)\
(=\Delta(0,j)+\Delta(0,j,j,l)+\Delta(0,j,l,n-1))\ge 0$, there are
only three possibilities for the values of  $\Delta(0,j)$ and
$\Delta(0,j,l,n-1))$: $\Delta(0,j)=0$ and $\Delta(0,j,l,n-1)> 0$;
$\Delta(0,j)>0$ and $\Delta(0,j,l,n-1)=0$; $\Delta(0,j)>0$ and
$\Delta(0,j,l,n-1)>0$. We consider each of these cases in turn and
show that, for all of them, we have a contradiction with the
choice of $0$ and $n-1$.
\begin{itemize}
\item[] \underline{Subcase 1.1} $\Delta(0,j)=0$ and $\Delta(0,j,l,n-1)>
0$\\
In addition to the previously stated precedence constraints
$0\prec j$, $l\prec j$, and $n-1\prec j$, we also have $l \prec
n-1$ (since $\Delta(0,j,l,n-1)>0$) and $l\prec 0$ (since
$\Delta(0,j,0,l)=\Delta(0,j)+\Delta(0,j,j,l)<0$). This gives two
possible permutations for permuting the submatrix
$M[\{0,j,l,n-1\}]$ to an a-Monge matrix: $\seq{l,0,n-1,j}$ and
$\seq{l,n-1,0,j}$. If the matrix obtained after permuting rows and
columns of $M[\{0,j,l,n-1\}]$ according to $\seq{l,0,n-1,j}$ is
indeed an a-Monge matrix, then the value $\Delta(l,j)$ can be
represented as a sum of non-negative numbers (see
equality~(\ref{eq:delta})) which include, in particular,
$\Delta(0,n-1)$ and $\Delta(0,j,l,0)=-\Delta(0,j,0,l)>0$. This
means that $\Delta(l,j)>\Delta(0,n-1)$, which is a contradiction
with the choice of $0$ and $n-1$. For the second permutation, the
value $\Delta(l,j)$ is represented as a sum of non-negative
numbers which include $\Delta(n-1,0) (=\Delta(0,n-1))$ and
$\Delta(0,j,l,n-1)>0$. This also yields
$\Delta(l,j)>\Delta(0,n-1)$.

\item[]  \underline{Subcase 1.2} $\Delta(0,j)>0$ and $\Delta(0,j,l,n-1)=
0$\\
In this subcase, we have $\Delta(j,n-1,l,n-1)\ge 0$ (since
$\Delta(0,n-1,l,n-1)=\Delta(0,j,l,n-1)+\Delta(j,n-1,l,n-1) \ge
0$), and therefore
$\Delta(j,n-1)=\Delta(j,n-1,j,l)+\Delta(j,n-1,l,n-1)>0$ (since
$\Delta(j,n-1,j,l)>0$) . The initial precedence constraints
$\{0\prec j, l\prec j, n-1\prec j\}$ imply that, in a-Monge
permutation for $M[\{0,j,l,n-1\}]$, we must have either $0\prec
n-1 \prec j$ or $n-1\prec 0\prec j$. In the former case, the value
$\Delta(0,j)$ can be represented as a sum of non-negative numbers
including $\Delta(0,n-1)$ and $\Delta(n-1,j)>0$ (see
equality~(\ref{eq:delta})), and hence $\Delta(0,j)>\Delta(0,n-1)$
which is a contradiction with the choice of $0$ and $n-1$. In the
latter case, we similarly get $\Delta(n-1,j)\ge
\Delta(n-1,0)+\Delta(0,j)$ and, since $\Delta(0,j)>0$ by the
assumption of Subcase 1.2, we obtain
$\Delta(n-1,j)>\Delta(0,n-1)$. Therefore, it follows that
$M[\{0,j,l,n-1\}]$ is a bad matrix.

\item[]  \underline{Subcase 1.3} $\Delta(0,j)>0$ and $\Delta(0,j,l,n-1)> 0$\\
There are three potential a-Monge permutations for
$M[\{0,j,l,n-1\}]$:  $\seq{0,l,n-1,j}$,
 $\seq{l,0,n-1,j}$, and  $\seq{l,n-1,0,j}$. As in the previous subcase,
 one can show that if one of them is an a-Monge permutation for
$M[\{0,j,l,n-1\}]$ then there is a contradiction with the choice of
$0$ and $n-1$.
\end{itemize}

\item[] \underline{Case 2} For any triple $i,j,l$, $i<j<l$,
with $\Delta(i,j,j,l)<0$, neither $i=0$ nor $l=n-1$.

It is easy to see that this condition implies the following
inequalities:
\begin{itemize}
\item[] $\Delta(0,i,j,l)>0$ because, otherwise,
$\Delta(0,j,j,l)=\Delta(0,i,j,l)+\Delta(i,j,j,l)<0$; \item[]
$\Delta(i,j,l,n-1)>0$ because, otherwise,
$\Delta(i,j,j,n-1)=\Delta(i,j,j,l)+\Delta(i,j,l,n-1)<0$; \item[]
$\Delta(j,l,l,n-1)\geq 0$; \item[] $\Delta(0,i,i,j)\geq 0$.

\end{itemize}

We claim that, given the above inequalities, at least one of
$M[\{0,i,j,l\}]$ and $M[\{i,j,l,n-1\}]$ is a bad matrix.

We show first that if $\Delta(j,l)= 0$, then $M[\{0,i,j,l\}]$ is
the bad matrix. Indeed, when trying to find an a-Monge permutation
for this matrix, the assumption $0\prec i$ yields $j\prec l$
(since $\Delta(0,i,j,l)>0$) and $i\prec l$ (since
$\Delta(0,i,i,l)=\Delta(0,i,i,j)+\Delta(0,i,j,l)>0$). The
constraint $j\prec l$ for the columns yields the constraint
$j\prec i$, and, since
$\Delta(i,l,j,l)=\Delta(i,j,j,l)+\Delta(j,l)=\Delta(i,j,j,l)<0$,
it also yields $l \prec i$. The contradictory precedence
constraints $\{l\prec i,i\prec l\}$ prove that $M[\{0,i,j,l\}]$ is
a bad matrix. So, we assume now that
 $\Delta(j,l)>0$ (the case $\Delta(j,l)<0$ is already eliminated).

By using a similar argument for the matrix $M[\{i,j,l,n-1\}]$), we
see that that if $\Delta(i,j)=0$ then this matrix is bad. So we
will also assume that $\Delta(i,j)>0$.

We now consider the submatrix $M[\{0,i,j,l\}]$ and will try to
permute it into an a-Monge matrix. The assumption $0\prec i$
yields $j\prec l$, $i\prec l$,  $j\prec i$. Since $\Delta(i,j)>0$,
and so $\Delta(0,j,i,j)=\Delta(0,i,i,j)+\Delta(i,j,i,j)
>0$, we also have $j\prec 0$. This shows that a permutation other
than $\seq{j,0,i,l}$ cannot be an a-Monge permutation for
$M[\{0,i,j,l\}]$. By analyzing the matrix $M[\{i,j,l,n-1\}]$ in a
similar way, we see that the only potential a-Monge permutation
for it is the permutation $\seq{i,l,n-1,j}$.

If $\seq{j,0,i,l}$ is an a-Monge permutation for $M[\{0,i,j,l\}]$
then  we must have $\Delta(0,i,j,0)\ge 0$. Since
$\Delta(0,i,i,j)\ge 0$ by the assumption of Case 2, and also
$\Delta(0,i)\ge 0$, we have
$\Delta(0,i,0,j)=\Delta(0,i,0,i)+\Delta(0,i,i,j)\ge 0$. However,
$\Delta(0,i,0,j)=-\Delta(0,i,j,0)$, which implies that
$\Delta(0,i)=0$ and $\Delta(0,i,i,j)=0$.

Moreover, if $\seq{j,0,i,l}$ is an a-Monge permutation for
$M[\{0,i,j,l\}]$ then $\Delta(j,i,i,l)\ge 0$. Similarly, if
$\seq{i,l,n-1,j}$ is an a-Monge permutation for $M[\{i,j,l,n-1\}]$
then $\Delta(i,j,i,l)\ge 0$. But
$\Delta(j,i,i,l)=-\Delta(i,j,i,l)$, so both are equal to 0.

If $\seq{j,0,i,l}$ is an a-Monge permutation for $M[\{0,i,j,l\}]$
then  we must have $\Delta(j,0,i,l)\ge 0$. We can express
$\Delta(j,0,i,l)$ as $\Delta(j,0,i,l)= -\Delta(0,j,i,l)=
-(\Delta(0,i,i,j)+\Delta(0,i,j,l)+\Delta(i,j)+\Delta(i,j,j,l))$.
Since $\Delta(0,i,i,j)=0$ and
$\Delta(i,j)+\Delta(i,j,j,l)=\Delta(i,j,i,l)=0$, we get
$\Delta(0,i,j,l)=-\Delta(j,0,i,l)\le 0$. However, the inequality
$\Delta(0,i,j,l)>0$ is one of the four inequalities (see above)
directly implied by the assumption of Case 2. Hence, we get a
contradiction which proves that at least one of the matrices
$M[\{0,i,j,l\}]$ and $M[\{i,j,l,n-1\}]$ is a bad matrix.
\end{itemize}

\noindent This completes the proof of the theorem.

\end{proof}

\noindent Note that the bound $|B|\le 4$ in the above theorem is
tight. Indeed, it can be straightforwardly checked that the
following matrix is not permuted a-Monge, while any matrix
obtained from it by deleting a row and a column (with the same
index) is permuted a-Monge.

\[ \left(%
\begin{array}{cccc}
  1 & 1 & 0 & 1\\
  1 & 1 & 0 & 0\\
  0 & 0 & 0 & 0 \\
  1 & 0 & 0 & 1
\end{array}%
\right)
\]

\noindent We can now derive the main result of this section from
Theorem~\ref{genmongerest}.

\begin{corollary}\label{01mongerest}
Let $M_1,\ldots,M_m$ be $n \times n$ matrices. If there exists no
permutation that simultaneously permutes all these matrices into
a-Monge matrices, then there exists a subset of indices $B$ with
 $|B|\leq 4$, such that no permutation of the
indices in $B$ simultaneously permutes matrices
$M_1[B],\ldots,M_m[B]$ into a-Monge matrices.
\end{corollary}
\begin{proof}
Consider the matrix $M=\sum_{i=1}^m{M_i}$. If $M$ is not a
permuted a-Monge matrix then, by Theorem~\ref{genmongerest}, there
exists a subset of indices $B$ with $|B| \leq 4$, such that $M[B]$
is a bad matrix. Consider matrices $M_1[B],\ldots,M_m[B]$. If
there existed a permutation that permutes all these matrices into
a-Monge matrices, then the sum of the permuted matrices, which is
$M[B]$, would be an a-Monge matrix as well. This contradiction
proves that there exists no permutation that simultaneously
permutes matrices $M_1[B],\ldots,M_m[B]$ into a-Monge matrices.

Assume now that $M$ is a permuted a-Monge matrix. The
corresponding a-Monge permutation does not permute
$M_1,\ldots,M_m$ into a-Monge matrices. Hence, there exist indices
$i,j,k,l$ and a pair of matrices, say, $M_1$ and $M_2$ such that
$M_1(i,k)+M_1(j,l)-M_1(i,l)-M_1(j,k)>0$ and
$M_2(i,k)+M_2(j,l)-M_2(i,l)-M_2(j,k)<0$. This implies that the
matrices $M_1[\{i,j,k,l\}]$ and $M_2[\{i,j,k,l\}]$ cannot be
simultaneously permuted into a-Monge matrices -- this follows from
the fact that the assumption $i \prec j$ implies $k \prec l$ for
$M_1$ and $l\prec k$ for $M_2$.
\end{proof}

\subsection{Reducing the number of matrices} \label{redF}

We will now prove the bound $|\F'|\le 3$ in
Proposition~\ref{bigred}, again via a-Monge matrices. In the
proof, we will use special partial orders which we call {\em
multipartite} partial orders.

We say that a partial order $\preceq$ on a set $D$ is multipartite
if and only if there is a partition of $D=D_1\cup\ldots\cup D_t$,
$t\ge 2$, such that $d\preceq d'$ if and only if  $d=d'$ or else
$d\in D_i$ and $d'\in D_j$ for some $1\le i<j\le t$. 
If $P$ is a multipartite order, then we will call the
classes $D_1,\ldots,D_t$ the corresponding {\em partition classes}
of $P$.

It is clear that if $\pi$ is an a-Monge permutation, for a matrix
$M$ then the reverse permutation $\pi^-$ is also an a-Monge
permutation for $M$. It is also clear that if $M$ is a-Monge and
the matrix obtained from $M$ by simultaneously swapping rows $s$
and $t$ and columns $s$ and $t$ is again a-Monge then rows $s$ and
$t$ are equivalent, i.e., $M(s,i)=M(t,i)+\alpha_{st}$, and columns
$s$ and $t$ are equivalent as well. Note that swapping of
equivalent rows and columns does not affect the property of being
a-Monge. A matrix $M$ is called Monge if $-M$ is a-Monge. It is
shown in Observation~3.6 of~\cite{Rudolf94:recognition} that if
$M$ is Monge,  $i\prec j\prec k$ in $M$, and rows (columns) $i,k$
are equivalent in $M$ then row $j$ is equivalent to these rows
(columns). Clearly, the statement is also true for a-Monge
matrices. Theorem~3.9 of~\cite{Rudolf94:recognition} states that
if a Monge matrix has no equivalent rows or columns then the only
way to permute it to a Monge matrix is by using either the
identity permutation $id$ or its reverse $id^-$.

This leads to the following characterization of a-Monge
permutations in terms of multipartite orders.  For every
anti-Monge square matrix $M$, there exists two mutually reverse
multipartite orders such that a permutation (i.e. ordering) of the
indices of $M$ is an a-Monge permutation if and only if this
ordering is an extension of one of the two multipartite orders.
Two indices $i,j$ belong to the same partition class of such a
multipartite order if and only if both rows $i,j$ and columns
$i,j$ are equivalent in $M$.

We will now prove two auxiliary lemmas about multipartite orders.

\begin{lemma}\label{compar}
For any two multipartite orders $P'$ and $P''$ on $D$, there are
$a,b\in D$ such that $a$ and $b$ are comparable (not necessarily
in the same direction) both in $P'$ and in $P''$.
\end{lemma}

\begin{proof}
Take a maximal chain in $P'$. If it is not entirely contained in a
class of $P''$ then there are two elements in this chain belonging
to two different classes of $P''$, that is, these elements are
comparable both in $P''$ and in $P'$. If all elements in the
maximal chain are contained in the same class of $P''$, then pick
any element $d$ in a different class of $P''$. This element is
comparable, in $P''$, with all elements from the chain, and,
clearly, it is comparable with at least one of these elements in
$P'$.
\end{proof}

Let us say that a collection $\mathcal{P}=\{P_1\zd P_l\}$ of
multipartite orders is {\em conflicting} if their union
(considered as a digraph $G_{\mathcal{P}}$) contains a directed
cycle.

\begin{lemma}\label{twoarcs}
If a collection $\mathcal{P}=\{P_1\zd P_l\}$ is conflicting then
the digraph $G_{\mathcal{P}}$ contains arcs $(a,b)$ and $(b,a)$
for some distinct $a,b$.
\end{lemma}

\begin{proof}
Let $a_1\zd a_t,a_1$ be a shortest directed cycle in $G$, and
assume, for contradiction, that $t>2$. Without loss of generality,
let $(a_1,a_2)\in P_1$. In this case, $(a_2,a_3)\not\in P_1$,
since, otherwise, we would have $(a_1,a_3)\in P_1$ and get a
shorter cycle. Without loss of generality, assume that
$(a_2,a_3)\in P_2$. Since the order $P_1$ is multipartite, we
conclude that $a_1$ and $a_3$ are comparable in $P_1$.
Furthermore, since we cannot have $(a_1,a_3)\in P_1$, we have
$(a_3,a_1)\in P_1$. Since $(a_1,a_2)\in P_1$, the transitivity of
$P_1$ implies that $(a_3,a_2)\in P_1$, which, together with
$(a_2,a_3)\in P_2$, gives us the required arcs.
\end{proof}

\begin{proposition}\label{redto3}
Let $U=\{M_1,\ldots,M_m\}$ be a set of matrices of size $n\times n$
such that no permutation is an a-Monge permutation for all matrices
in $U$. Then, there is a subset $U'\sse U$ such that $|U'|\le 3$ and
no permutation is an a-Monge permutation for all matrices in $U'$.
\end{proposition}

\begin{proof}
We may assume that every matrix in $U$ is a permuted a-Monge
matrix, since, otherwise, the result follows immediately. Start
with matrix $M_1\in U$ and choose any of the two multipartite
orders that describe the set of corresponding a-Monge permutations
for $M_1$. Call this order $P_1$. By Lemma~\ref{compar}, there is
a pair $(a,b)\in P_1$ such that $a\ne b$ and  $a$ and $b$  are
comparable in $P_2$, where $P_2$ is the multipartite order for
$M_2$. We may assume that $(a,b)\in P_2$, since, otherwise, the
other multipartite order for $P_2$ would be chosen.

If there is a pair of distinct elements $(c,d)$ such that
$(c,d)\in P_1$ and $(d,c)\in P_2$, then there exists no a-Monge
permutation for $M_1$ and $M_2$ and the proposition is proved. So
we may assume that $\{P_1,P_2\}$ is not conflicting. Since $P_1$
shares a pair of comparable elements with any multipartite order,
we can in the same way choose a multipartite order $P_i$ for each
matrix $M_i$. If, for some $i$, the pair $\{P_1,P_i\}$ is
conflicting, then the proposition is proved. So assume that all
such pairs of orders are non-conflicting. Note that if we chose
the other multipartite order for $M_1$, this would have led to
choosing the other multipartite orders for all $M_1\zd M_m$.

Since there is no common a-Monge permutation for all of $M_1\zd
M_m$, we know that the collection $\{P_1\zd P_m\}$ of orders that
we have constructed is conflicting. By Lemma~\ref{twoarcs}, there
are orders $P_i$ and $P_j$ such that, for some distinct $e,f$, we
have $(e,f)\in P_i$ and $(f,e)\in P_j$. Since both $P_i$ and $P_j$
share with $P_1$ some pairs of elements comparable in the same
direction, we conclude that there is no common a-Monge permutation
for $M_1,M_i,M_j$. This completes the proof.
\end{proof}

\noindent Note that the bound $|U'|\le 3$ in the above proposition
is tight. Indeed, each of the following three matrices is permuted
a-Monge, every two of them have a common a-Monge permutation, but
there is no common a-Monge permutation for all three of them.

\[ \left(%
\begin{array}{ccc}
  1 & 0 & 0 \\
  0 & 0 & 0 \\
  0 & 0 & 0 \\
\end{array}%
\right) \; \;
\left(%
\begin{array}{ccc}
  0 & 0 & 0 \\
  0 & 1 & 0 \\
  0 & 0 & 0 \\
\end{array}%
\right) \; \;
\left(%
\begin{array}{ccc}
  0 & 0 & 0 \\
  0 & 0 & 0 \\
  0 & 0 & 1 \\
\end{array}%
\right)
\]

\section*{Appendix B: Strict implementations from Case 1}

It is assumed throughout that $D=\{0,1,2,3\}$. Implementations
should be read as follows:
\begin{itemize}
\item the symbol
$\stackrel{s\phantom{8pt}}{\Longrightarrow_\alpha}$ means
``strictly $\alpha$-implements'';

\item ${\cal U}$ always denotes ${\cal U}_D$;
\item $Y=\{x,y\}$ is the set of primary variables and $Z=\{z,w\}$
is the set of auxiliary variables (see
Definition~\ref{strictdef}).
\end{itemize}

Each implementation produces some predicate $g$ such that either
$g$ or $\pi(g)$, or $\pi(g^c)$ (for some permutation $\pi$) is a
predicate for which a strict implementation has already been
found, or else a predicate $g$ such that, for some $D'\subset D$,
$g|_{D'}$ is not supermodular on any chain on $D'$. We will
describe the latter situation by writing, for simplicity, that
``$g|_{D'}$ is bad''. If $|D'|=2$ then one can directly verify
that the corresponding matrix is not a-Monge (there is no need to
permute rows and columns). For the case $|D'|=3$, one can use
Lemma~\ref{mongestru} to quickly check that the matrix of
$g|_{D'}$ is not a permuted a-Monge matrix.

{\flushleft
{\small 1. $\left\{h'_{1} := {\footnotesize \begin{array}{l}
1000\\
0110\\
1000\\
0000\\
\end{array}
}\right\}$ $\cup$ ${\cal U}
\stackrel{s\phantom{8pt}}{\Longrightarrow_2}$ ${\footnotesize
\begin{array}{l}
1000\\
1110\\
1000\\
0000\\
\end{array}
} =: g$ \hskip 0.5cm
$g | _{\{0,1,3\}}$ is bad\\
{\small $g(x,y) + 1 = max_{z}[h'_{1}(z,y)+h'_{1}(x,z)+u_{\{3\}}(z)]$}\\
}\vskip 0.4cm

{\small 2. $\left\{h'_{2} := {\footnotesize \begin{array}{l}
1000\\
1101\\
1000\\
0000\\
\end{array}
}\right\}$ $\cup$ ${\cal U}
\stackrel{s\phantom{8pt}}{\Longrightarrow_3}$ ${\footnotesize
\begin{array}{l}
1000\\
1101\\
1010\\
0000\\
\end{array}
} =: g$ \hskip 0.5cm
$g | _{\{0,2,3\}}$  is bad\\
{\small $g(x,y) + 2 = max_{z}[h'_{2}(z,x)+h'_{2}(z,y)+h'_{2}(x,y)+u_{\{3\}}(z)+u_{\{2\}}(x)+u_{\{2\}}(y)]$}\\
}\vskip 0.4cm

{\small 3. $\left\{h'_{3} := {\footnotesize \begin{array}{l}
1001\\
0111\\
1110\\
1001\\
\end{array}
}\right\}$ $\cup$ ${\cal U}
\stackrel{s\phantom{8pt}}{\Longrightarrow_4}$ ${\footnotesize
\begin{array}{l}
1000\\
0111\\
1110\\
0001\\
\end{array}
} =: g$ \hskip 0.5cm
$g | _{\{0,1,3\}}$ is bad\\
{\small $g(x,y) + 3 = max_{z}[h'_{3}(z,x)+h'_{3}(z,y)+h'_{3}(x,y)+u_{\{1,2\}}(z)]$}\\
}\vskip 0.4cm

{\small 4. $\left\{h'_{4} := {\footnotesize \begin{array}{l}
1010\\
0101\\
1010\\
1000\\
\end{array}
}\right\}$ $\cup$ ${\cal U}
\stackrel{s\phantom{8pt}}{\Longrightarrow_4}$ ${\footnotesize
\begin{array}{l}
1101\\
0101\\
1000\\
0101\\
\end{array}
} =: g$ \hskip 0.5cm
$g | _{\{0,1,2\}}$ is bad\\
{\small $g(x,y) + 3 = max_{z,w}[h'_{4}(z,w)+h'_{4}(z,y)+h'_{4}(w,x)+u_{\{1,3\}}(z)]$}\\
}\vskip 0.4cm

{\small 5. $\left\{h'_{5} := {\footnotesize \begin{array}{l}
1010\\
0110\\
0000\\
0000\\
\end{array}
}\right\}$ $\cup$ ${\cal U}
\stackrel{s\phantom{8pt}}{\Longrightarrow_3}$ ${\footnotesize
\begin{array}{l}
1000\\
0100\\
0001\\
0001\\
\end{array}
} =: g$ \hskip 0.5cm
$g | _{\{0,1,3\}}$ is bad\\
{\small $g(x,y) + 2 = max_{z}[h'_{5}(x,z)+h'_{5}(x,y)+h'_{5}(y,z)+u_{\{3\}}(z)+u_{\{2,3\}}(x)+u_{\{3\}}(y)]$}\\
}\vskip 0.4cm

{\small 6. $\left\{h'_{6} := {\footnotesize \begin{array}{l}
1010\\
0111\\
1010\\
1000\\
\end{array}
}\right\}$ $\cup$ ${\cal U}
\stackrel{s\phantom{8pt}}{\Longrightarrow_4}$ ${\footnotesize
\begin{array}{l}
1010\\
0111\\
1010\\
1001\\
\end{array}
} =: g$ \hskip 0.5cm
$g | _{\{0,1,3\}}$ is bad\\
{\small $g(x,y) + 3 = max_{z}[h'_{6}(x,z)+h'_{6}(x,y)+h'_{6}(y,z)+u_{\{2\}}(z)+u_{\{3\}}(x)+u_{\{3\}}(y)]$}\\
}\vskip 0.4cm

{\small 7. $\left\{h'_{7} := {\footnotesize \begin{array}{l}
1010\\
0111\\
1110\\
1000\\
\end{array}
}\right\}$ $\cup$ ${\cal U}
\stackrel{s\phantom{8pt}}{\Longrightarrow_3}$ ${\footnotesize
\begin{array}{l}
1110\\
0000\\
0000\\
1010\\
\end{array}
} =: g$ \hskip 0.5cm
$g | _{\{2,3\}}$ is bad\\
{\small $g(x,y) + 2 = max_{z}[h'_{7}(z,y)+h'_{7}(x,z)+u_{\{0,3\}}(x)]$}\\
}\vskip 0.4cm

{\small 8. $\left\{h'_{8} := {\footnotesize \begin{array}{l}
1011\\
0101\\
1010\\
0000\\
\end{array}
}\right\}$ $\cup$ ${\cal U}
\stackrel{s\phantom{8pt}}{\Longrightarrow_6}$ ${\footnotesize
\begin{array}{l}
1010\\
0000\\
1010\\
1011\\
\end{array}
} =: g$ \hskip 0.5cm
$g | _{\{0,1,3\}}$ is bad\\
{\small $g(x,y) + 5 = max_{z,w}[h'_{8}(z,w)+h'_{8}(z,x)+h'_{8}(z,y)+h'_{8}(w,x)+u_{\{2\}}(z)+u_{\{0\}}(w)+u_{\{1,3\}}(x)]$}\\
}\vskip 0.4cm

{\small 9. $\left\{h'_{9} := {\footnotesize \begin{array}{l}
1011\\
0111\\
0010\\
0000\\
\end{array}
}\right\}$ $\cup$ ${\cal U}
\stackrel{s\phantom{8pt}}{\Longrightarrow_3}$ ${\footnotesize
\begin{array}{l}
1001\\
0101\\
0010\\
1101\\
\end{array}
} =: g$ \hskip 0.5cm
$g | _{\{0,1,2\}}$ is bad\\
{\small $g(x,y) + 2 = max_{z}[h'_{9}(z,x)+h'_{9}(x,y)+h'_{9}(y,z)+u_{\{3\}}(z)+u_{\{3\}}(x)+u_{\{3\}}(y)]$}\\
}\vskip 0.4cm

{\small 10. $\left\{h'_{10} := {\footnotesize \begin{array}{l}
1011\\
0111\\
0010\\
0001\\
\end{array}
}\right\}$ $\cup$ ${\cal U}
\stackrel{s\phantom{8pt}}{\Longrightarrow_3}$ ${\footnotesize
\begin{array}{l}
1011\\
0111\\
0010\\
0000\\
\end{array}
} =: g$ \hskip 0.5cm
$g = h'_{9}$\\
{\small $g(x,y) + 2 = max_{z}[h'_{10}(z,x)+h'_{10}(z,y)+u_{\{2\}}(z)+u_{\{0,1\}}(x)]$}\\
}\vskip 0.4cm

{\small 11. $\left\{h'_{11} := {\footnotesize \begin{array}{l}
1011\\
0111\\
0011\\
0000\\
\end{array}
}\right\}$ $\cup$ ${\cal U}
\stackrel{s\phantom{8pt}}{\Longrightarrow_2}$ ${\footnotesize
\begin{array}{l}
1000\\
0100\\
0000\\
1100\\
\end{array}
} =: g$ \hskip 0.5cm
$\pi(g^t) = h'_{5} \mbox{ where } \pi(0,1,2,3)=(0,1,3,2)$\\
{\small $g(x,y) + 1 = h'_{11}(x,y)+u_{\{3\}}(x)+u_{\{0,1\}}(y)$}\\
}\vskip 0.4cm

{\small 12. $\left\{h'_{12} := {\footnotesize \begin{array}{l}
1011\\
0111\\
0110\\
1001\\
\end{array}
}\right\}$ $\cup$ ${\cal U}
\stackrel{s\phantom{8pt}}{\Longrightarrow_3}$ ${\footnotesize
\begin{array}{l}
0110\\
0111\\
0111\\
0000\\
\end{array}
} =: g$ \hskip 0.5cm
$g | _{\{0,1,3\}}$ is bad\\
{\small $g(x,y) + 2 = max_{z}[h'_{12}(z,y)+h'_{12}(x,z)+u_{\{1,2\}}(z)]$}\\
}\vskip 0.4cm

{\small 13. $\left\{h'_{13} := {\footnotesize \begin{array}{l}
1011\\
0111\\
1010\\
0000\\
\end{array}
}\right\}$ $\cup$ ${\cal U}
\stackrel{s\phantom{8pt}}{\Longrightarrow_3}$ ${\footnotesize
\begin{array}{l}
1011\\
0101\\
1010\\
0000\\
\end{array}
} =: g$ \hskip 0.5cm
$g = h'_{8}$\\
{\small $g(x,y) + 2 = max_{z}[h'_{13}(z,x)+h'_{13}(x,y)+h'_{13}(y,z)+u_{\{3\}}(z)+u_{\{3\}}(y)]$}\\
}\vskip 0.4cm

{\small 14. $\left\{h'_{14} := {\footnotesize \begin{array}{l}
1011\\
0111\\
1010\\
0001\\
\end{array}
}\right\}$ $\cup$ ${\cal U}
\stackrel{s\phantom{8pt}}{\Longrightarrow_3}$ ${\footnotesize
\begin{array}{l}
0001\\
0111\\
0000\\
0001\\
\end{array}
} =: g$ \hskip 0.5cm
$\pi(g) = h'_{2} \mbox{ where } \pi(0,1,2,3)=(3,1,0,2)$\\
{\small $g(x,y) + 2 = max_{z}[h'_{14}(z,y)+h'_{14}(x,z)+u_{\{1,3\}}(z)]$}\\
}\vskip 0.4cm

{\small 15. $\left\{h'_{15} := {\footnotesize \begin{array}{l}
1011\\
0111\\
1110\\
0000\\
\end{array}
}\right\}$ $\cup$ ${\cal U}
\stackrel{s\phantom{8pt}}{\Longrightarrow_4}$ ${\footnotesize
\begin{array}{l}
1011\\
0101\\
1010\\
0000\\
\end{array}
} =: g$ \hskip 0.5cm
$g = h'_{8}$\\
{\small $g(x,y) + 3 = max_{z}[h'_{15}(x,z)+h'_{15}(x,y)+h'_{15}(y,z)+u_{\{0,3\}}(z)+u_{\{3\}}(x)+u_{\{3\}}(y)]$}\\
}\vskip 0.4cm

{\small 16. $\left\{h'_{16} := {\footnotesize \begin{array}{l}
1011\\
0111\\
1110\\
0001\\
\end{array}
}\right\}$ $\cup$ ${\cal U}
\stackrel{s\phantom{8pt}}{\Longrightarrow_4}$ ${\footnotesize
\begin{array}{l}
1011\\
0000\\
1011\\
0001\\
\end{array}
} =: g$ \hskip 0.5cm
$g | _{\{0,1,3\}}$  is bad\\
{\small $g(x,y) + 3 = max_{z}[h'_{16}(z,x)+h'_{16}(z,y)+h'_{16}(x,z)+u_{\{0,3\}}(z)]$}\\
}\vskip 0.4cm

{\small 17. $\left\{h'_{17} := {\footnotesize \begin{array}{l}
1011\\
0111\\
1110\\
1001\\
\end{array}
}\right\}$ $\cup$ ${\cal U}
\stackrel{s\phantom{8pt}}{\Longrightarrow_4}$ ${\footnotesize
\begin{array}{l}
1010\\
0111\\
1110\\
0001\\
\end{array}
} =: g$ \hskip 0.5cm
$g | _{\{0,1,3\}}$ is bad\\
{\small $g(x,y) + 3 = max_{z}[h'_{17}(z,x)+h'_{17}(z,y)+h'_{17}(x,y)+u_{\{1,2\}}(z)]$}\\
}\vskip 0.4cm

{\small 18. $\left\{h'_{18} := {\footnotesize \begin{array}{l}
1011\\
0111\\
1110\\
1101\\
\end{array}
}\right\}$ $\cup$ ${\cal U}
\stackrel{s\phantom{8pt}}{\Longrightarrow_5}$ ${\footnotesize
\begin{array}{l}
1110\\
0000\\
1110\\
1110\\
\end{array}
} =: g$ \hskip 0.5cm
$g | _{\{1,3\}}$  is bad\\
{\small $g(x,y) + 4 = max_{z,w}[h'_{18}(z,w)+h'_{18}(z,y)+h'_{18}(w,x)+u_{\{1,2\}}(z)+u_{\{0\}}(w)]$}\\
}\vskip 0.4cm

{\small 19. $\left\{h'_{19} := {\footnotesize \begin{array}{l}
1011\\
1101\\
1010\\
0000\\
\end{array}
}\right\}$ $\cup$ ${\cal U}
\stackrel{s\phantom{8pt}}{\Longrightarrow_5}$ ${\footnotesize
\begin{array}{l}
1010\\
1011\\
1010\\
1010\\
\end{array}
} =: g$ \hskip 0.5cm
$g | _{\{1,3\}}$  is bad\\
{\small $g(x,y) + 4 = max_{z,w}[h'_{19}(z,w)+h'_{19}(z,y)+h'_{19}(x,z)+u_{\{2\}}(z)+u_{\{2\}}(w)+u_{\{1,3\}}(x)]$}\\
}\vskip 0.4cm

{\small 20. $\left\{h'_{20} := {\footnotesize \begin{array}{l}
1100\\
1101\\
1000\\
0000\\
\end{array}
}\right\}$ $\cup$ ${\cal U}
\stackrel{s\phantom{8pt}}{\Longrightarrow_2}$ ${\footnotesize
\begin{array}{l}
1110\\
1101\\
1011\\
0111\\
\end{array}
} =: g$ \hskip 0.5cm
$\pi(g)= h'_{18} \mbox{ where } \pi(0,1,2,3)=(0,3,1,2)$\\
{\small $g(x,y) + 1 = h'_{20}(x,y)+h'_{20}(y,x)+u_{\{2,3\}}(x)+u_{\{2,3\}}(y)$}\\
}\vskip 0.4cm

{\small 21. $\left\{h'_{21} := {\footnotesize \begin{array}{l}
1101\\
0110\\
0110\\
1001\\
\end{array}
}\right\}$ $\cup$ ${\cal U}
\stackrel{s\phantom{8pt}}{\Longrightarrow_6}$ ${\footnotesize
\begin{array}{l}
1001\\
1101\\
0000\\
1001\\
\end{array}
} =: g$ \hskip 0.5cm
$g | _{\{0,1,2\}}$  is bad\\
{\small $g(x,y) + 5 = max_{z,w}[h'_{21}(z,w)+h'_{21}(z,x)+h'_{21}(z,y)+h'_{21}(w,x)+u_{\{3\}}(z)+u_{\{0\}}(w)+u_{\{1,2\}}(x)]$}\\
}\vskip 0.4cm

{\small 22. $\left\{h'_{22} := {\footnotesize \begin{array}{l}
1101\\
1100\\
0010\\
0000\\
\end{array}
}\right\}$ $\cup$ ${\cal U}
\stackrel{s\phantom{8pt}}{\Longrightarrow_3}$ ${\footnotesize
\begin{array}{l}
1000\\
1110\\
0010\\
1010\\
\end{array}
} =: g$ \hskip 0.5cm
$\pi(g^t)= h'_{9} \mbox{ where } \pi(0,1,2,3)=(0,2,1,3)$\\
{\small $g(x,y) + 2 = max_{z}[h'_{22}(x,z)+h'_{22}(y,z)+u_{\{2,3\}}(z)+u_{\{1,3\}}(x)]$}\\
}\vskip 0.4cm

{\small 23. $\left\{h'_{23} := {\footnotesize \begin{array}{l}
1101\\
1110\\
0000\\
0000\\
\end{array}
}\right\}$ $\cup$ ${\cal U}
\stackrel{s\phantom{8pt}}{\Longrightarrow_2}$ ${\footnotesize
\begin{array}{l}
1101\\
1110\\
0111\\
1011\\
\end{array}
} =: g$ \hskip 0.5cm
$\pi(g) = h'_{18} \mbox{ where } \pi(0,1,2,3)=(0,2,1,3)$\\
{\small $g(x,y) + 1 = h'_{23}(x,y)+h'_{23}(y,x)+u_{\{2,3\}}(x)+u_{\{2,3\}}(y)$}\\
}\vskip 0.4cm

{\small 24. $\left\{h'_{24} := {\footnotesize \begin{array}{l}
1101\\
1110\\
0110\\
1001\\
\end{array}
}\right\}$ $\cup$ ${\cal U}
\stackrel{s\phantom{8pt}}{\Longrightarrow_6}$ ${\footnotesize
\begin{array}{l}
1001\\
1101\\
0000\\
1001\\
\end{array}
} =: g$ \hskip 0.5cm
$g | _{\{0,1,2\}}$ is bad\\
{\small $g(x,y) + 5 = max_{z,w}[h'_{24}(z,w)+h'_{24}(z,x)+h'_{24}(z,y)+h'_{24}(x,z)+u_{\{3\}}(z)+u_{\{3\}}(w)+u_{\{1,2\}}(x)]$}\\
}\vskip 0.4cm

{\small 25. $\left\{h'_{25} := {\footnotesize \begin{array}{l}
1110\\
1100\\
0000\\
0000\\
\end{array}
}\right\}$ $\cup$ ${\cal U}
\stackrel{s\phantom{8pt}}{\Longrightarrow_2}$ ${\footnotesize
\begin{array}{l}
0000\\
1101\\
0001\\
0001\\
\end{array}
} =: g$ \hskip 0.5cm
$\pi(g^t) = h'_{2} \mbox{ where } \pi(0,1,2,3)=(1,3,0,2)$\\
{\small $g(x,y) + 1 = h'_{25}(x,y)+u_{\{1,2,3\}}(x)+u_{\{3\}}(y)$}\\
}\vskip 0.4cm

{\small 26. $\left\{h'_{26} := {\footnotesize \begin{array}{l}
1110\\
1100\\
1010\\
0000\\
\end{array}
}\right\}$ $\cup$ ${\cal U}
\stackrel{s\phantom{8pt}}{\Longrightarrow_2}$ ${\footnotesize
\begin{array}{l}
0000\\
1101\\
1011\\
0001\\
\end{array}
} =: g$ \hskip 0.5cm
$\pi(g) = h'_{9} \mbox{ where } \pi(0,1,2,3)=(1,2,3,0)$\\
{\small $g(x,y) + 1 = h'_{26}(x,y)+u_{\{1,2,3\}}(x)+u_{\{3\}}(y)$}\\
}\vskip 0.4cm

{\small 27. $\left\{h'_{27} := {\footnotesize \begin{array}{l}
1110\\
1101\\
1010\\
0000\\
\end{array}
}\right\}$ $\cup$ ${\cal U}
\stackrel{s\phantom{8pt}}{\Longrightarrow_2}$ ${\footnotesize
\begin{array}{l}
0110\\
0101\\
0010\\
0111\\
\end{array}
} =: g$ \hskip 0.5cm
$\pi(g^t) = h'_{13} \mbox{ where } \pi(0,1,2,3)=(1,2,3,0)$\\
{\small $g(x,y) + 1 = h'_{27}(x,y)+u_{\{3\}}(x)+u_{\{1,2,3\}}(y)$}\\
}\vskip 0.4cm
}

\section*{Appendix C: Strict implementations from Case 2}

The rules for reading implementations are the same as in Appendix
B. Each implementation implements some predicate $g$ such that,
for some $D'\subset D$, $g|_{D'}$ is bad, or else a pair for which
a strict implementation has already been found.

{\flushleft
{\small 1. $\left\{h := {\footnotesize \begin{array}{l}
1100\\
0000\\
0000\\
0001\\
\end{array}
},f := {\footnotesize \begin{array}{l}
1000\\
0001\\
0000\\
0001\\
\end{array}
}\right\} \cup$ ${\cal U}
\stackrel{s\phantom{8pt}}{\Longrightarrow_{3}}$ ${\footnotesize
\begin{array}{l}
1000\\
1101\\
0000\\
0101\\
\end{array}
} =: g\qquad$ $g | _{\{0,1,2\}}$ is bad}
\\
{\small $g(x,y) + 2 =
max_{z}[f(x,z)+f(y,z)+h(z,x)+u_{\{1\}}(z)+u_{\{1,2\}}(x)]$}

\vskip 0.3cm

{\small 2. $\left\{h := {\footnotesize \begin{array}{l}
1100\\
0000\\
0000\\
0001\\
\end{array}
},f := {\footnotesize \begin{array}{l}
1110\\
0001\\
0000\\
0001\\
\end{array}
}\right\} \cup$ ${\cal U}
\stackrel{s\phantom{8pt}}{\Longrightarrow_{3}}$ ${\footnotesize
\begin{array}{l}
1000\\
0001\\
0000\\
0001\\
\end{array}
} =: g\qquad$ $(h,g)$ is Pair 1}
\\
{\small $g(x,y) + 2 =
max_{z}[f(x,z)+h(x,z)+h(y,z)+u_{\{2\}}(z)+u_{\{1,2\}}(x)]$}

\vskip 0.3cm

{\small 3. $\left\{h := {\footnotesize \begin{array}{l}
1100\\
0000\\
0000\\
0001\\
\end{array}
},f := {\footnotesize \begin{array}{l}
1000\\
1001\\
1000\\
0001\\
\end{array}
}\right\} \cup$ ${\cal U}
\stackrel{s\phantom{8pt}}{\Longrightarrow_{4}}$ ${\footnotesize
\begin{array}{l}
0101\\
0000\\
0000\\
0101\\
\end{array}
} =: g\qquad$ $g | _{\{0,1\}}$ is bad}
\\
{\small $g(x,y) + 3 =
max_{w,z}[f(z,w)+f(y,w)+h(x,z)+u_{\{2\}}(z)+u_{\{3\}}(w)]$}

\vskip 0.3cm

{\small 4. $\left\{h := {\footnotesize \begin{array}{l}
1100\\
0000\\
0000\\
0001\\
\end{array}
},f := {\footnotesize \begin{array}{l}
1010\\
1010\\
1010\\
0001\\
\end{array}
}\right\} \cup$ ${\cal U}
\stackrel{s\phantom{8pt}}{\Longrightarrow_{2}}$ ${\footnotesize
\begin{array}{l}
1000\\
1001\\
1000\\
0001\\
\end{array}
} =: g\qquad$ $(h,g)$ is Pair 3}
\\
{\small $g(x,y) + 1 = max_{z}[f(z,x)+h(y,z)+u_{\{1\}}(x)]$}

\vskip 0.3cm

{\small 5. $\left\{h := {\footnotesize \begin{array}{l}
1100\\
0000\\
0000\\
0001\\
\end{array}
},f := {\footnotesize \begin{array}{l}
1010\\
1011\\
1010\\
0001\\
\end{array}
}\right\} \cup$ ${\cal U}
\stackrel{s\phantom{8pt}}{\Longrightarrow_{2}}$ ${\footnotesize
\begin{array}{l}
1000\\
1001\\
1000\\
0001\\
\end{array}
} =: g\qquad$ $(h,g)$ is Pair 3}
\\
{\small $g(x,y) + 1 = max_{z}[f(x,z)+h(y,z)]$}

\vskip 0.3cm

{\small 6. $\left\{h := {\footnotesize \begin{array}{l}
1100\\
0000\\
0000\\
0001\\
\end{array}
},f := {\footnotesize \begin{array}{l}
1110\\
0001\\
1110\\
0001\\
\end{array}
}\right\} \cup$ ${\cal U}
\stackrel{s\phantom{8pt}}{\Longrightarrow_{3}}$ ${\footnotesize
\begin{array}{l}
1010\\
1010\\
1010\\
0001\\
\end{array}
} =: g\qquad$ $(h,g)$ is Pair 4}
\\
{\small $g(x,y) + 2 = max_{z}[f(z,x)+f(z,y)+f(y,z)]$}

\vskip 0.3cm

{\small 7. $\left\{h := {\footnotesize \begin{array}{l}
1100\\
0000\\
0000\\
0001\\
\end{array}
},f := {\footnotesize \begin{array}{l}
1010\\
1011\\
1011\\
0001\\
\end{array}
}\right\} \cup$ ${\cal U}
\stackrel{s\phantom{8pt}}{\Longrightarrow_{2}}$ ${\footnotesize
\begin{array}{l}
1000\\
1001\\
1000\\
0001\\
\end{array}
} =: g\qquad$ $(h,g)$ is Pair 3}
\\
{\small $g(x,y) + 1 = f(y,x)+u_{\{1\}}(x)+u_{\{0,3\}}(y)$}

\vskip 0.3cm

{\small 8. $\left\{h := {\footnotesize \begin{array}{l}
1110\\
0000\\
0001\\
0001\\
\end{array}
},f := {\footnotesize \begin{array}{l}
1010\\
0001\\
0001\\
0001\\
\end{array}
}\right\} \cup$ ${\cal U}
\stackrel{s\phantom{8pt}}{\Longrightarrow_{3}}$ ${\footnotesize
\begin{array}{l}
1000\\
1101\\
0101\\
0101\\
\end{array}
} =: g\qquad$ $g | _{\{0,1,2\}}$ is bad}
\\
{\small $g(x,y) + 2 =
max_{z}[f(z,y)+f(y,z)+h(x,z)+u_{\{1\}}(x)+u_{\{1\}}(y)]$}

\vskip 0.3cm

{\small 9. $\left\{h := {\footnotesize \begin{array}{l}
1110\\
0000\\
0001\\
0001\\
\end{array}
},f := {\footnotesize \begin{array}{l}
1000\\
0001\\
1001\\
0001\\
\end{array}
}\right\} \cup$ ${\cal U}
\stackrel{s\phantom{8pt}}{\Longrightarrow_{2}}$ ${\footnotesize
\begin{array}{l}
1010\\
0000\\
0111\\
0111\\
\end{array}
} =: g\qquad$ $g | _{\{0,1,2\}}$ is bad}
\\
{\small $g(x,y) + 1 = max_{z}[f(y,z)+h(x,z)]$}

\vskip 0.3cm

{\small 10. $\left\{h := {\footnotesize \begin{array}{l}
1110\\
0000\\
0001\\
0001\\
\end{array}
},f := {\footnotesize \begin{array}{l}
1010\\
0101\\
0101\\
0101\\
\end{array}
}\right\} \cup$ ${\cal U}
\stackrel{s\phantom{8pt}}{\Longrightarrow_{4}}$ ${\footnotesize
\begin{array}{l}
1010\\
0001\\
0001\\
0001\\
\end{array}
} =: g\qquad$ $(h,g)$ is Pair 8}
\\
{\small $g(x,y) + 3 = max_{z}[f(z,y)+f(x,z)+h(z,y)+u_{\{0,3\}}(z)]$}

\vskip 0.3cm

{\small 11. $\left\{h := {\footnotesize \begin{array}{l}
1110\\
0000\\
0001\\
0001\\
\end{array}
},f := {\footnotesize \begin{array}{l}
1000\\
0101\\
1101\\
0101\\
\end{array}
}\right\} \cup$ ${\cal U}
\stackrel{s\phantom{8pt}}{\Longrightarrow_{3}}$ ${\footnotesize
\begin{array}{l}
1010\\
0000\\
0111\\
0111\\
\end{array}
} =: g\qquad$ $g | _{\{0,1,2\}}$ is bad}
\\
{\small $g(x,y) + 2 = max_{z}[f(y,z)+h(x,z)+u_{\{0,3\}}(z)]$}

\vskip 0.3cm

{\small 12. $\left\{h := {\footnotesize \begin{array}{l}
1110\\
0000\\
0001\\
0001\\
\end{array}
},f := {\footnotesize \begin{array}{l}
1000\\
1101\\
1101\\
0101\\
\end{array}
}\right\} \cup$ ${\cal U}
\stackrel{s\phantom{8pt}}{\Longrightarrow_{2}}$ ${\footnotesize
\begin{array}{l}
1000\\
0001\\
1001\\
0001\\
\end{array}
} =: g\qquad$ $(h,g)$ is Pair 9}
\\
{\small $g(x,y) + 1 = f(y,x)+u_{\{2\}}(x)+u_{\{0,3\}}(y)$}

\vskip 0.3cm

{\small 13. $\left\{h := {\footnotesize \begin{array}{l}
1000\\
1001\\
0001\\
0001\\
\end{array}
},f := {\footnotesize \begin{array}{l}
1000\\
1001\\
1000\\
0001\\
\end{array}
}\right\} \cup$ ${\cal U}
\stackrel{s\phantom{8pt}}{\Longrightarrow_{4}}$ ${\footnotesize
\begin{array}{l}
1000\\
1101\\
0101\\
0101\\
\end{array}
} =: g\qquad$ $g | _{\{0,1,2\}}$ is bad}
\\
{\small $g(x,y) + 3 =
max_{z}[f(z,y)+f(y,z)+h(x,z)+u_{\{3\}}(z)+u_{\{0,1,2\}}(y)]$}

\vskip 0.3cm

{\small 14. $\left\{h := {\footnotesize \begin{array}{l}
1000\\
1001\\
0001\\
0001\\
\end{array}
},f := {\footnotesize \begin{array}{l}
1010\\
1011\\
1010\\
0001\\
\end{array}
}\right\} \cup$ ${\cal U}
\stackrel{s\phantom{8pt}}{\Longrightarrow_{3}}$ ${\footnotesize
\begin{array}{l}
1000\\
1001\\
1000\\
0001\\
\end{array}
} =: g\qquad$ $(h,g)$ is Pair 13}
\\
{\small $g(x,y) + 2 = max_{z}[f(z,y)+f(x,z)+h(y,z)]$}

\vskip 0.3cm

{\small 15. $\left\{h := {\footnotesize \begin{array}{l}
1000\\
1001\\
0001\\
0001\\
\end{array}
},f := {\footnotesize \begin{array}{l}
1010\\
1011\\
1011\\
0001\\
\end{array}
}\right\} \cup$ ${\cal U}
\stackrel{s\phantom{8pt}}{\Longrightarrow_{2}}$ ${\footnotesize
\begin{array}{l}
1000\\
1001\\
1000\\
0001\\
\end{array}
} =: g\qquad$ $(h,g)$ is Pair 13}
\\
{\small $g(x,y) + 1 = f(y,x)+u_{\{1\}}(x)+u_{\{0,3\}}(y)$}

\vskip 0.3cm

{\small 16. $\left\{h := {\footnotesize \begin{array}{l}
1100\\
1101\\
0001\\
0001\\
\end{array}
},f := {\footnotesize \begin{array}{l}
1010\\
1101\\
1010\\
0000\\
\end{array}
}\right\} \cup$ ${\cal U}
\stackrel{s\phantom{8pt}}{\Longrightarrow_{3}}$ ${\footnotesize
\begin{array}{l}
1100\\
0101\\
1011\\
0101\\
\end{array}
} =: g\qquad$ $g | _{\{0,1,2\}}$ is bad}
\\
{\small $g(x,y) + 2 = max_{z}[f(y,x)+h(z,y)+h(x,z)+u_{\{3\}}(z)]$}

\vskip 0.3cm

{\small 17. $\left\{h := {\footnotesize \begin{array}{l}
1100\\
1101\\
0001\\
0001\\
\end{array}
},f := {\footnotesize \begin{array}{l}
1100\\
1100\\
1011\\
0000\\
\end{array}
}\right\} \cup$ ${\cal U}
\stackrel{s\phantom{8pt}}{\Longrightarrow_{4}}$ ${\footnotesize
\begin{array}{l}
1100\\
0100\\
0011\\
0111\\
\end{array}
} =: g\qquad$ $g | _{\{0,1,2\}}$ is bad}
\\
{\small $g(x,y) + 3 =
max_{z}[f(x,y)+h(x,z)+h(y,z)+u_{\{3\}}(z)+u_{\{0,3\}}(x)]$}

\vskip 0.3cm

{\small 18. $\left\{h := {\footnotesize \begin{array}{l}
1100\\
1101\\
0001\\
0001\\
\end{array}
},f := {\footnotesize \begin{array}{l}
0000\\
1101\\
1011\\
0000\\
\end{array}
}\right\} \cup$ ${\cal U}
\stackrel{s\phantom{8pt}}{\Longrightarrow_{4}}$ ${\footnotesize
\begin{array}{l}
0000\\
1011\\
1011\\
1011\\
\end{array}
} =: g\qquad$ $g | _{\{0,1\}}$ is bad}
\\
{\small $g(x,y) + 3 = max_{w,z}[f(w,y)+h(w,z)+h(x,z)+u_{\{2\}}(w)]$}

\vskip 0.3cm

{\small 19. $\left\{h := {\footnotesize \begin{array}{l}
1100\\
1101\\
0001\\
0001\\
\end{array}
},f := {\footnotesize \begin{array}{l}
1010\\
0001\\
1011\\
0001\\
\end{array}
}\right\} \cup$ ${\cal U}
\stackrel{s\phantom{8pt}}{\Longrightarrow_{2}}$ ${\footnotesize
\begin{array}{l}
1010\\
1111\\
0111\\
0111\\
\end{array}
} =: g\qquad$ $g | _{\{0,1,2\}}$ is bad}
\\
{\small $g(x,y) + 1 = max_{z}[f(y,z)+h(x,z)]$}

\vskip 0.3cm

{\small 20. $\left\{h := {\footnotesize \begin{array}{l}
1100\\
1101\\
0001\\
0001\\
\end{array}
},f := {\footnotesize \begin{array}{l}
1110\\
0101\\
0000\\
0101\\
\end{array}
}\right\} \cup$ ${\cal U}
\stackrel{s\phantom{8pt}}{\Longrightarrow_{2}}$ ${\footnotesize
\begin{array}{l}
1010\\
0001\\
1011\\
0001\\
\end{array}
} =: g\qquad$ $(h,g)$ is Pair 19}
\\
{\small $g(x,y) + 1 = f(x,y)+u_{\{2\}}(x)+u_{\{0,2,3\}}(y)$}

\vskip 0.3cm

{\small 21. $\left\{h := {\footnotesize \begin{array}{l}
1100\\
1101\\
0001\\
0001\\
\end{array}
},f := {\footnotesize \begin{array}{l}
1010\\
0101\\
1010\\
0101\\
\end{array}
}\right\} \cup$ ${\cal U}
\stackrel{s\phantom{8pt}}{\Longrightarrow_{4}}$ ${\footnotesize
\begin{array}{l}
1100\\
0111\\
0000\\
0111\\
\end{array}
} =: g\qquad$ $g | _{\{0,1,2\}}$ is bad}
\\
{\small $g(x,y) + 3 = max_{z}[f(z,x)+h(x,z)+h(y,z)+u_{\{0,3\}}(z)]$}

\vskip 0.3cm

{\small 22. $\left\{h := {\footnotesize \begin{array}{l}
1100\\
1101\\
0001\\
0001\\
\end{array}
},f := {\footnotesize \begin{array}{l}
0000\\
0101\\
1011\\
0101\\
\end{array}
}\right\} \cup$ ${\cal U}
\stackrel{s\phantom{8pt}}{\Longrightarrow_{2}}$ ${\footnotesize
\begin{array}{l}
1010\\
1101\\
1010\\
0000\\
\end{array}
} =: g\qquad$ $(h,g)$ is Pair 16}
\\
{\small $g(x,y) + 1 = f(y,x)+u_{\{0,1,2\}}(x)+u_{\{0\}}(y)$}

\vskip 0.3cm

{\small 23. $\left\{h := {\footnotesize \begin{array}{l}
1100\\
1101\\
0001\\
0001\\
\end{array}
},f := {\footnotesize \begin{array}{l}
0000\\
1101\\
0011\\
0011\\
\end{array}
}\right\} \cup$ ${\cal U}
\stackrel{s\phantom{8pt}}{\Longrightarrow_{2}}$ ${\footnotesize
\begin{array}{l}
1100\\
1100\\
1011\\
0000\\
\end{array}
} =: g\qquad$ $(h,g)$ is Pair 17}
\\
{\small $g(x,y) + 1 = f(y,x)+u_{\{0,1,2\}}(x)+u_{\{0\}}(y)$}

\vskip 0.3cm

{\small 24. $\left\{h := {\footnotesize \begin{array}{l}
1000\\
1001\\
1001\\
0001\\
\end{array}
},f := {\footnotesize \begin{array}{l}
0000\\
1101\\
1011\\
0000\\
\end{array}
}\right\} \cup$ ${\cal U}
\stackrel{s\phantom{8pt}}{\Longrightarrow_{4}}$ ${\footnotesize
\begin{array}{l}
0000\\
1011\\
1011\\
1011\\
\end{array}
} =: g\qquad$ $g | _{\{0,1\}}$ is bad}
\\
{\small $g(x,y) + 3 =
max_{w,z}[f(z,w)+f(w,y)+h(x,z)+u_{\{1,3\}}(z)+u_{\{2\}}(w)]$}

\vskip 0.3cm

{\small 25. $\left\{h := {\footnotesize \begin{array}{l}
1100\\
1100\\
1101\\
0001\\
\end{array}
},f := {\footnotesize \begin{array}{l}
1001\\
0100\\
1101\\
1001\\
\end{array}
}\right\} \cup$ ${\cal U}
\stackrel{s\phantom{8pt}}{\Longrightarrow_{4}}$ ${\footnotesize
\begin{array}{l}
1101\\
1001\\
1101\\
1101\\
\end{array}
} =: g\qquad$ $g | _{\{0,1\}}$ is bad}
\\
{\small $g(x,y) + 3 =
max_{w,z}[f(z,y)+f(x,w)+h(z,w)+u_{\{0\}}(z)+u_{\{3\}}(w)]$}

\vskip 0.3cm

{\small 26. $\left\{h := {\footnotesize \begin{array}{l}
1100\\
1100\\
1101\\
0001\\
\end{array}
},f := {\footnotesize \begin{array}{l}
1101\\
0100\\
1101\\
1001\\
\end{array}
}\right\} \cup$ ${\cal U}
\stackrel{s\phantom{8pt}}{\Longrightarrow_{3}}$ ${\footnotesize
\begin{array}{l}
1001\\
0100\\
1101\\
1001\\
\end{array}
} =: g\qquad$ $(h,g)$ is Pair 25}
\\
{\small $g(x,y) + 2 =
max_{z}[f(z,x)+f(z,y)+u_{\{1,3\}}(z)+u_{\{2\}}(x)]$}

\vskip 0.3cm

{\small 27. $\left\{h := {\footnotesize \begin{array}{l}
1100\\
1100\\
0011\\
0011\\
\end{array}
},f := {\footnotesize \begin{array}{l}
1001\\
0110\\
0110\\
1001\\
\end{array}
}\right\} \cup$ ${\cal U}
\stackrel{s\phantom{8pt}}{\Longrightarrow_{4}}$ ${\footnotesize
\begin{array}{l}
1111\\
1001\\
1001\\
1111\\
\end{array}
} =: g\qquad$ $g | _{\{0,1\}}$ is bad}
\\
{\small $g(x,y) + 3 =
max_{w,z}[f(z,y)+f(w,x)+h(z,w)+u_{\{3\}}(z)+u_{\{0\}}(w)]$}

\vskip 0.3cm
}

\end{document}